\documentstyle[12pt,psfig,draft,array,float,amssymb]{nature-ejb-newr}

\def\simlt{\mathrel{\hbox{\rlap{\hbox{\lower4pt\hbox{$\sim$}}}\hbox{$<$}}}}
\def\simgt{\mathrel{\hbox{\rlap{\hbox{\lower4pt\hbox{$\sim$}}}\hbox{$>$}}}}

\def\ale{\mathrel{\hbox{\rlap{\hbox{\lower4pt\hbox{$\sim$}}}\hbox{$<$}}}}
\def\age{\mathrel{\hbox{\rlap{\hbox{\lower4pt\hbox{$\sim$}}}\hbox{$>$}}}}

\def\gtorder{\mathrel{\raise.3ex\hbox{$>$}\mkern-14mu
             \lower0.6ex\hbox{$\sim$}}}
\def\ltorder{\mathrel{\raise.3ex\hbox{$<$}\mkern-14mu
             \lower0.6ex\hbox{$\sim$}}}

\def\citebraketlow#1{${}$#1${}$}%
\def\citebraket#1{$^{}$#1$^{}$}%
\def\citebraketd#1#2{$^{}$#1$^{,}$#2$^{}$}%

\def\spose#1{\hbox to 0pt{#1\hss}}
\newcommand\lsim{\mathrel{\spose{\lower 3pt\hbox{$\mathchar"218$}}
     \raise 2.0pt\hbox{$\mathchar"13C$}}}
\newcommand\gsim{\mathrel{\spose{\lower 3pt\hbox{$\mathchar"218$}}
     \raise 2.0pt\hbox{$\mathchar"13E$}}}
     
\newcommand{\Lbol}{$L_{\mathrm bol}$}
\newcommand{\Mbol}{$M_{\mathrm bol}$}
\newcommand{\Msol}{M$_{\odot}$}
\newcommand{\Rsol}{R$_{\odot}$}
\newcommand{\Lx}{$L_{\mathrm X}$}
\newcommand{\Mx}{$M_{\mathrm BH}$}
\newcommand{\Mv}{$M_{\mathrm V}$}
\newcommand{\Mopt}{$M_{\mathrm *}$}
\newcommand{\chisqr}{$\chi^2$}
\newcommand{\cnts}{cts\,s$^{-1}$}
\newcommand{\Porb}{$P_{\mathrm{orb}}$}
\newcommand{\Ps}{$P_{\mathrm{s}}$}
\newcommand{\degr}{$^{\circ}$}
\newcommand\arcsec{\mbox{$^{\prime\prime}$}}%
\newcommand\arcmin{\mbox{$^\prime$}}%
\newcommand{\ergscm}{\,erg\,cm$^{-2}$\,s$^{-1}$}
\newcommand{\kms}{km\,s$^{-1}$}

\newcommand{\Hbet}{H${\beta}$}
\newcommand{\Teff}{$T_{\mathrm eff}$}
\newcommand{\maxvel}{$\approx$\,160}
\newcommand{\ergs}{\,erg\,s$^{-1}$}
\newcommand{\Av}{$A_{\mathrm V}$}

\def\farcs{\hbox{$.\!\!^{\prime\prime}$}}
\newcounter{iref}
\begin{document} 

\title{A mass of less than 15 solar masses for the black hole in an ultraluminous X-ray source}

\author{
C. Motch\affiliation[1]{Observatoire astronomique de Strasbourg, Universit\'e de Strasbourg, CNRS, UMR 7550, 11 rue de l'Universit\'e, F-67000 Strasbourg, France}, M. W. Pakull$^{1}$,
R. Soria\affiliation[2]{International Centre for Radio Astronomy Research, Curtin University, GPO Box U1987, Perth, WA 6845, Australia},
F. Gris\'e$^{1,}$\affiliation[3]{Instituto de Astrofisica de Canarias C/ V\'ia L\'actea, s/n, E38205 - La Laguna (Tenerife), Spain.}$^{,}$\affiliation[4]{Universidad de La Laguna, Dept. Astrofísica, E-38206 La Laguna, Tenerife, Spain,}$^{,}$\affiliation[5]{Department of Physics and Astronomy, University of Iowa, Van Allen Hall, Iowa City, IA 52242, USA},
G. Pietrzy\'nski\affiliation[6]{Universidad de Concepci\'on, Departamento de Astronom\'{\i}a, Casilla 160-C, Concepci\'on, Chile}$^{,}$\affiliation[7]{Warsaw University Observatory, Al. Ujazdowskie 4, 00-478 Warszawa, Poland.}
}
      
\date{\today}{}
\headertitle{}
\mainauthor{Motch et al.}

\summary{

Most ultraluminous X-ray sources\citebraket{\cite{feng2011}} (ULXs) display a typical set of properties not seen in Galactic stellar-mass black holes (BHs): higher luminosity ($L_{\rm{x}} > 3 \times 10^{39}$\,\ergs), unusually soft X-ray components ($kT \lesssim 0.3$ keV) and a characteristic downturn\citebraketd{\cite{gladstone2009}}{\cite{roberts2005}} in their spectra above $\approx 5$ keV. Such puzzling properties have been interpreted either as evidence of intermediate-mass BHs\citebraketd{\cite{colbert1999}}{\cite{miller2003}}, or as emission from stellar-mass BHs accreting above their Eddington limit\citebraketd{\cite{poutanen2007}}{\cite{kawashima2012}}, analogous to some Galactic BHs at peak luminosity\citebraketd{\cite{done2004}}{\cite{middleton2013}}. Recently, a very soft X-ray spectrum has been observed in a rare and transient stellar-mass BH\citebraket{\cite{liu2013}}. Here we show that the X-ray source P13 in the galaxy NGC\,7793\citebraket{\cite{motch2011}} is in a $\approx$\,64\,day period binary and exhibits all three canonical properties of ULXs. By modelling the strong optical and UV modulations due to X-ray heating of the B9Ia donor star, we constrain the BH mass to less than 15 solar masses. Our results demonstrate that in P13, soft thermal emission and spectral curvature are indeed signatures of supercritical accretion. By analogy, ULXs with similar X-ray spectra and luminosities of up to a few 10$^{40}$\,\ergs\ can be explained by supercritical accretion onto massive stellar BHs.
}
\maketitle


We organised an X-ray, UV and optical spectrophotometric monitoring programme of ULX P13\citebraket{\cite{read1999}} from late 2009 until late 2013 using {\it Swift, Chandra, XMM-Newton} and the ESO-VLTs.  These data were supplemented by photometry obtained at the Warsaw 1.3\,m telescope (Las Campanas Observatory) in 2004 and 2005 and by an archival {\it Chandra} observation obtained in 2003. 

{\it Chandra} detected P13 in 2003\citebraket{\cite{panuti2011}} with a $0.3$--$10$\,keV X-ray luminosity of $\approx$\,4\,$\times$\,10$^{39}$\,\ergs. Our 73\,d long {\it Swift} X-ray Telescope (XRT) monitoring carried out in 2010 recorded a $0.3$--$10$\,keV X-ray luminosity in the range of 4.8\,$\pm$\,0.5\,$\times$\,10$^{39}$\,\ergs\ down to less than 1.6\,$\times$\,10$^{38}$\,\ergs\ at one occasion. A similar X-ray luminosity of 2.0\,$\pm$\,0.1\,$\times$\,10$^{39}$\,\ergs\ was detected in our last {\it XMM-Newton} observation in November 2013. The 2003 {\it Chandra} spectrum displays a spectral break at $\approx$\,4.2\,keV. Our 2013 {\it XMM-Newton} observation confirms the break seen by {\it Chandra} and reveals a soft  disk blackbody-like component with  $kT_{\mathrm in}$\,$\approx$\,0.3 keV (Extended Data Tab.~1 \& Fig.~\ref{figext:ChandraXMMSpec}). Therefore, P13 exhibits all the hallmarks of a canonical ultraluminous X-ray state\citebraketd{\cite{gladstone2009}}{\cite{roberts2005}}.   
Remarkably, {\it Swift}/XRT did not detect P13 in any of the individual pointings performed from August 2011 until June 2013. Stacking {\it Swift} low state data reveals the source at \Lx\,($0.3$--$10$\,keV)\,=\,5\,$\pm$\,1\,$\times$\,10$^{37}$\,\ergs\ (90\% confidence level), a factor 100 less than in the previously seen bright X-ray state. Scheduled and serendipitous {\it Chandra} and {\it XMM-Newton} observations carried out in 2011 and 2012 detected the source at the same low X-ray luminosity and will be reported elsewhere. 

Optical spectra point at a B9I spectral type (Fig.~1). In addition to high-order Balmer absorption lines, the spectrum exhibits Balmer emission up to at least H$\gamma$ as well as HeII $\lambda$4686 and Bowen CIII-NIII complex in emission. Assuming that minimum light ($V$\,=\,20.50\,mag; Extended Data Figs \ref{figext:opticaldata}{\it abcd} \& \ref{figext:Swift}{\it a}) represents stellar light only yields an absolute magnitude of \Mv\,=\,-7.50 ($d$\,=\,3.7\,Mpc\citebraket{\cite{radburn2011}}, mean \Av\,=\,0.16\citebraketd{\cite{schlegel1998}}{\cite{piet2010}}). Such a high optical luminosity implies a type Ia supergiant classification with initial masses of 20\,-\,25\,\Msol and present masses of 18\,-\,23\,\Msol\citebraket{\cite{meynet2000}}. With \Lbol\ $\approx$ 5\,$\times$\,10$^{38}$\,\ergs, almost one tenth of the maximum observed X-ray luminosity we expect X-ray heating effects to be noticeable. 
  
The Las Campanas observations and the {\it Swift}/UVOT {\it u} and $V$ band 2010 monitorings both show two consecutive maxima providing possible hints of a $\approx$ 2 month long orbital period. The power spectrum analysis of the 2004\,-\,2011 $V$ light-curve (Extended Data Fig.~\ref{figext:power_Vheii}) reveals two aliasing periods at $P$\,=\,65.165\,d and at $P$\,=\,63.340\,d. Corresponding periodicities are found in the HeII radial velocities. Importantly, analysing times of $V$ and UV photometric maxima over a 8 year time interval reveals a jitter of up to $\pm$\,0.09 in phase which might reflect a superorbital period of $\sim$ 5 to 8.8\,yr (Extended Data Figs \ref{figext:UVOTmaxima} \& \ref{figext:PeriodFitResiduals}). 

The pattern of radial velocity variations with orbital phase changed very significantly between 2010 (X-ray bright state) and 2011 (X-ray faint state) (Extended Data Fig.~\ref{figext:foldedHeII}). Balmer absorption lines show what may be a coherent variation with orbital phase in 2010 while the pattern of variability is clearly more complex in 2011. The total velocity amplitude of the absorption lines is \maxvel\,\kms.  

Interestingly, the shape and amplitude of the $V$ optical light-curve (Extended Data \ref{figext:foldedLC}) and the mean value of the HeII $\lambda$4686 equivalent width (EW) do not seem to depend on the observed X-ray luminosity. The relative amplitude of the {\it u} ($\lambda_{\mathrm central} $\,=\,3465\AA;  $\Delta\,u$\,=\,1.0\,mag) and $V$ ($\lambda_{\mathrm central} $\,=\,5500\AA; $\Delta$\,$V$\,=\,0.5 \,mag) light-curves and the behaviour of the $V$--$I$ colour index point at a strong X-ray heating effect of the supergiant star hemisphere facing the compact companion. An X-ray source with a tenth of the nominal X-ray luminosity would brighten the star by only $\approx$ 0.1 magnitude in $V$ at maximum light and would have basically no effect in the faint X-ray state. Therefore, we conclude that in 2011 part of the companion star photosphere continues to be illuminated by a luminous X-ray source which is however shielded from our view. The Galactic X-ray binary Her~X-1 exhibits similar bright/faint X-ray states\citebraket{\cite{giacconi1973}} as well as periodic phase shifts of photometric maxima\citebraket{\cite{deeter1976}}. By analogy, we suggest that a tilted precessing accretion disk is at the origin of both the X-ray bright \& faint states and of the phase jitter of optical maximum light. 

In order to constrain the geometry and dynamics of the system we simultaneously fitted the $V$ and UVOT {\it u} light-curves using the Eclipsing Light Curve code\citebraket{\cite{orosz2000}} (ELC). We tested four X-ray luminosity levels ranging from 0.7 up to 2 times a nominal value of 4.2\,$\times$\,10$^{39}$\,\ergs\ (derived from the {\it diskbb + comptt} fit to the {\it Chandra} spectrum extrapolated to the $0.3$--$20$\,keV range), in order to account for the observed X-ray variability and for a possible undetected component radiating at energies below or above the observed $0.3$--$10$\,keV X-ray range. The ELC model included a dark accretion disk casting X-ray shadows on the X-ray heated stellar hemisphere. Modelling the optical disk emission caused by X-ray heating as well as the absence of veiling in the high order Balmer absorption lines both indicate that optical light is fully dominated by stellar emission (Extended Data Fig.\ref{figext:EWcomparison}). Examples of fitted light-curves are shown in Fig.~2.

We explored the parameter space spanning mass ratios \Mx/\Mopt\ from 0.1 to 10, inclinations from 0 to 90\degr, 3 values of \Teff\,(11,000\,$\pm$\,1,000\,K), four choices of \Lx\ mentioned above, B9Ia star masses of 18 and 23\,\Msol, and two rotation states of the mass-donor star, namely synchronised at periastron and non rotating. In each case, the B9Ia star was assumed to fill its Roche lobe at periastron since the total wind mass loss rate of a B9Ia star\citebraket{\cite{kudritzki1999}} is lower or equal to the mass accretion rate required to explain the bright X-ray state ($\approx$ 7\,$\times$\,10$^{-7}$\,\Msol\,yr$^{-1}$). The accretion disk had a fixed radius of 0.7 times the Roche lobe radius at periastron. For each of these parameters, we obtained best fit values for the remaining orbital and disk parameters. We computed \Mbol\ by equating the radius of the mass-donor star with that of the corresponding volume averaged Roche lobe at periastron. We then constrained the range of possible masses of the X-ray source by forcing i) ELC fits to be acceptable at the 99.7\% level, ii) computed bolometric magnitudes to be consistent with observations and iii) possible eclipses to be shorter than the maximum of $\approx$ 7\,days allowed from {\it Swift} 2010 data. All acceptable orbital solutions implied velocity amplitudes of Balmer absorption lines smaller than the maximum observed range. 

All constraints converge towards a black hole mass lower than $\approx$ 15\,\Msol\ irrespective of the incident X-ray luminosity in the range considered here (see Fig.~3). Black holes more massive than $\approx$ 15\,\Msol\ imply too small Roche lobes at periastron to accommodate the large B9Ia star. All acceptable orbital solutions require a significant eccentricity of $e\,=\,0.27\,-\,0.41$ (see Table 1). The full radial velocity amplitude of the compact object varies from 120 to 290\,\kms, a range of values consistent with those observed in 2010 and 2011 for the HeII emission line.

The evolved nature of the donor star suggests that mass transfer to the black hole happens on a thermal timescale ($\approx\,$10$^5$\,yrs) as the supergiant rapidly expands\citebraket{\cite{pod2003}}.
This is more than an order of magnitude shorter than its main sequence lifetime and implies that supergiant ULXs are much more rare than systems with unevolved mass-donors\citebraket{\cite{rap2005}}.
Given the significant eccentricity of the orbit, it is likely that Roche-lobe overflow started only after the star began crossing the Hertzsprung gap\citebraket{\cite{patruno2010}}.

The intrinsic X-ray luminosity of $\approx$ 4\,$\times$\,10$^{39}$\,\ergs\ is about twice the Eddington luminosity of a 15\,\Msol\ accreting black hole. We thus confirm that P13 is a genuine Eddington or super-Eddington source and that its extreme X-ray luminosity does not reflect the presence of an intermediate-mass black hole.  Hence we do have direct evidence that the characteristic ULX X-ray spectrum with both a medium energy break and a soft X-ray excess is the signature of an Eddington or super-Eddington regime.

\medskip
\medskip

\noindent Supplementary Information is available in the online version of the paper.

\noindent 

\begin{acknowledge}
RS acknowledges an Australian Research Council's Discovery Projects funding scheme (project number DP 120102393). We thank J. Orosz for providing us with the ELC code. We gratefully acknowledge the effort of the {\it Swift} team for the execution of our observing programme. This work is based on observations made with ESO Telescopes at the La Silla Paranal Observatory under programmes ID 084.D-0881 and 087.D-0602 and uses observations made with the NASA/ESA Hubble Space Telescope, and obtained from the Hubble Legacy Archive, which is a collaboration between the Space Telescope Science Institute (STScI/NASA), the Space Telescope European Coordinating Facility (ST-ECF/ESA) and the Canadian Astronomy Data Centre (CADC/NRC/CSA). Support from the Ideas Plus program of Polish Ministry of Science
is also acknowledged. 
The scientific results reported in this article are based in part on observations made by the {\it Chandra} X-ray Observatory and in part on observations obtained with {\it XMM-Newton} (OBSIDs 0693760101 \& 0693760401), an ESA science mission with instruments and contributions directly funded by ESA Member States and NASA.
\end{acknowledge}

\bigskip
\noindent
{\bf Author contribution:}

CM wrote the manuscript with comments from all authors. MWP identified the optical counterpart of P13 and initiated optical observations at ESO. CM and FG analysed the spectroscopic and photometric data from several runs at the ESO VLT. FG analysed the HST data. GP provided photometric data from the Warsaw telescope at the Las Campanas observatory. RS, FG and CM designed and analysed the {\it Chandra}, {\it XMM-Newton} and {\it Swift} X-ray observation. CM carried out the light-curve fitting using the ELC code. CM, MP, RS  and FG made significant contributions to the interpretation and discussion of the data. All authors participated in the review of the manuscript.

\bigskip
\noindent

The authors declare no competing financial interests. Reprints and permissions information is available at
npg.nature.com/reprintsandpermissions.  Correspondence should be
addressed to C. Motch (e-mail: christian.motch@unistra.fr).

\clearpage

\medskip
\bigskip
\centerline{\bf Table 1}
\bigskip
\begin{center}
\medskip
\begin{tabular}{lcc}
\hline
Synchronised    &      Yes	    &        No \\
\hline
\Mx              &  3.45 - 6.9	    &    3.45 - 14.95\\
Min \chisqr     &  0.77 	    & 	   0.73 \\
$e$               & 0.27 - 0.38	    &   0.30 - 0.41\\
$\omega$        & 79  - 118         &     84 - 124\\
$i$               &  25  -  80	    &     20 - 80\\
\Mbol           & -7.72 -7.03	    &   -7.89 -6.97\\
$R_{\mathrm RL}$     &  64  - 77         &    64 - 82 \\
$X-ray\ heating$   &   0 - 0.04	    &     0  - 0.09\\
$RV2$             & 171 - 285	    &   124 - 286\\
$RV1$             &  18  - 64	    & 	   19 - 116 \\   
\hline  		  
\end{tabular}		  
\end{center}		  

\bigskip \noindent {\bf Table~1: Allowed range of orbital parameters.}

\noindent Caption: Min \chisqr\ lists the minimum reduced value (39 dof) of all acceptable solutions. $\omega$ is the periastron angle, $R_{\mathrm RL}$ is the radius of the Roche lobe of the mass donor star in \Rsol,  $X-ray\  heating$ is the additional $V$ band magnitude due to X-ray heating at minimum light. $RV1$ and $RV2$ are the full amplitudes (\kms) of the radial velocity of the mass-donor star and of the black hole respectively. All angles are in degrees.

\clearpage
\bigskip \noindent {\bf Figure~1: The mass donor star of P13 has a B9Ia spectral type.} 

\noindent {Caption: The rectified mean of 11 spectra obtained in 2010 and of 11 spectra obtained in 2011 for a total integration time of 15.52 hours and corresponding to $V$$_{\rm aver}$\,=\,20.31\,mag. All spectra were acquired at the ESO-VLT using FORS2. The MgII $\lambda$4481\AA\ absorption line is deeper than observed in 2009\citebraket{\cite{motch2011}} ($V$$_{\rm aver}$\,=\,20.20\,mag) and is now more consistent with a B9 type\citebraket{\cite{lennon1992}}. This indicates that the spectral type of P13 may vary slightly with X-ray illumination as expected. The strength of the SiII doublet at $\lambda$4128-4130\AA\ (panel {\it a)}) compared to neighbouring HeI lines also points at this spectral type. In a similar manner, the appearance of a group of FeII lines around $\lambda$4500\AA\ (panel {\it b)}) argues in favour of a B9Ia spectral type. A subset of spectra obtained close to minimum light ($V$$_{\rm aver}$\,=\,20.4\,mag) suggests the same spectral type, indicating that the intrinsic (non X-ray illuminated) spectral type of the mass donor star is indeed B9Ia. Temperature scales for B supergiants\citebraketd{\cite{markova2008}}{\cite{zorec2009}} imply an effective temperature of 11,000\,$\pm$\,1,000\,K. 

\clearpage
\bigskip \noindent {\bf Figure~2: Examples of $V$ and {\it u} light-curve fits}. 

\noindent Caption: We show here examples of acceptable models, using the ELC code, obtained by simultaneously fitting $V$ and {\it Swift}/UVOT {\it u} light-curves. Model parameters: mass-donor mass = 23\,\Msol, \Teff\,=\,12,000\,K, \Mx\,=\,6.9\,\Msol, Log(\Lx)\,=\,39.925, $i$ = 80\degr, $e$ = 0.33 and periastron angle = 93\degr. The disk radius has a $\beta$ opening angle of 6.9\degr. \chisqr\,=\,38.19 for 39 dof. Error bars in all panels show statistical error uncertainties at the 1$\sigma$ level. We assume an orbital period of 63.52\,d and a superorbital period of 2,620\,d. The spin angular velocity of the donor star is synchronized with its orbital angular velocity at periastron.

\noindent Panels {\it a)} and {\it b)}: ELC model fits to the {\it u} and binned $V$ light-curves. Occultation of the X-ray heated star hemisphere by the dark disk accounts for the small dip present at maximum light. Panel {\it c)}: the model radial velocity curve of the X-ray source overplotted on observed HeII emission line radial velocities. The model curve appears shifted by $\approx$\,0.3 in phase with respect to the observations. This suggests that in P13 the location of the HeII $\lambda$4686 emitting region does not accurately trace the motion of the compact object, a situation similar to that encountered in several Galactic low-mass X-ray binaries\citebraketd{\cite{still1997}}{\cite{pearson2006}}.    

\clearpage
\bigskip \noindent {\bf Figure~3: Constraints on the mass of the compact star}.

\noindent {Caption: Constraints on the mass of the compact object plotted for \Teff\,=\,11,000\,$\pm$\,1,000\,K (consistent with the B9Ia spectral type of the mass-donor star), Log(\Lx)\,=\,39.475, 39.625, 39.775 and 39.925, B9Ia stellar masses of 18 and 23\,\Msol, mass ratios \Mx/\Mopt in the range of 0.1 to 10 and mass donor stars non rotating (panel {\it a)}) or synchronized with orbital rotation at periastron (panel {\it b)}). For each mass-donor star mass, system inclination, mass ratio and \Teff, the best ELC fit to the {\it u} and $V$ light-curves provides the eccentricity, the periastron angle and phase, the radial velocity of the light barycentre of the companion star and of the X-ray source. We only show solutions with \Mx $>$ 3\,\Msol. Small black filled squares: solutions providing statistically acceptable fits to the light-curves at the 99.7\% confidence level. Large symbols: Solutions implying mass donor optical luminosities compatible with the size of the Roche lobe radius at periastron. Large green filled squares: excluded solutions implying eclipse durations longer than the maximum allowed by our {\it Swift} X-ray monitoring in 2010. Large red diamonds: finally allowed black hole masses considering all possible values of the input parameters. The maximum allowed black hole mass is $\approx$\,7\,\Msol\ and $\approx$\,15\,\Msol\ for a synchronized and a non rotating mass-donor star respectively. These values are reached for \Mopt\,= 23\,\Msol.
}
\newpage

\setcounter{figure}{0}

\noindent {\bf Extended Data Table 1.} {\bf X-ray Spectral Fits.} Spectral fits to {\it Chandra} (C03) and {\it XMM-Newton} (X13) bright state X-ray spectra, from 2003 and 2013 respectively. All errors are given at the 90\% confidence level. $P(H0)$ is the null hypothesis probability. In all cases, we assume a fixed Galactic column density $n_{\rm H} = 1.2 \times 10^{20}$ cm$^{-2}$ ({\it tbabs} model) in addition to a fitted intrinsic (NGC\,7793 + local) absorption column. For the {\it diskir} model, we assumed\citebraket{\cite{gierlinski2009} $f_{\mathrm in} = 0.1$, $r_{\mathrm irr} = 1.2$, $f_{\mathrm out} = 0.005$ and $\log r_{\mathrm out} = 5.0$.}

\begin{figure}[H]
\caption[]{{\bf XMM and Chandra bright state spectra.} The E\,$\times$\,f(E) unfolded energy distribution of the 2003 {\it Chandra}/ACIS-S (high, red datapoints) and of the 2013 {\it XMM-Newton}/EPIC combined spectra (low, blue datapoints) fitted with the {\it diskir} model. Best fit model parameters are listed in Extended Data Table 1. Spectra shown here have been rebinned with a minimum signal-to-noise ratio of 8 in each bin for display purposes. Error bars show the 1$\sigma$ statistical error.}
\end{figure}

\begin{figure}[H]
\caption[]{{\bf Optical Data.} {Optical light and radial velocity curves of P13. Panels {\it a)} \& {\it b)} show the 2004 and 2005 Las Campanas photometry. Other panels show the ESO-VLT light-curves ({\sl c) \& d)}), EW of the HeII\,$\lambda$4686 emission line ({\sl e) \& f))}, radial velocity curves of Balmer absorption lines ({\sl g) \& h)}) and of the HeII emission line ({\sl i) \& j)}). P13 was X-ray bright in 2010 (panels {\sl c), e), g) \& i)}) and X-ray faint in 2011 (panels {\sl d), f), h) \& j)}). Red arrows in panel {\it d)} mark times of the {\it Chandra} (near optical maximum) and {\it Swift} observations which detected P13 in the faint X-ray state. Error bars in all panels show statistical error uncertainties at the 1$\sigma$ level.}}
\end{figure}

\begin{figure}[H]
\caption{ {{\bf The 2010 observations.} Panel {\sl a)}: red squares; UVOT {\it u} ($\lambda_{\mathrm eff}$\,=\,3470\,\AA) band, black lozenges; ESO-VLT $V$ light-curve. Note the successive maxima separated by $\approx$\,64\,d, more pronounced in the {\it u} band ($\Delta$\,u\,$\approx$\,1.0\,mag) than in $V$ ($\Delta$\,V\,$\approx$\,0.5\,mag). Panel  {\sl b)}: {\it Swift}/XRT $0.3$--$10$\,keV light-curve obtained from 2010-08-16 until 2010-10-27. The count to flux factor was computed by fitting the {\it XMM-Newton} {\it diskir} model to the average bright state spectrum. This implies  unabsorbed ($0.3$--$10$\,keV) X-ray luminosities in the range of 4.8\,$\times$\,10$^{39}$\,\ergs\ down to less than 1.6\,$\times$\,10$^{38}$\,\ergs\ on MJD~55492.9 with a weighted average of (2.8\,$\pm$\,0.2)\,$\times$\,10$^{39}$\,\ergs. Error bars in all panels show statistical error uncertainties at the 1$\sigma$ level. Errors on $V$ magnitudes (not shown) are typically lower than 0.04 mag.}} 
\end{figure}

\begin{figure}[H]
\caption{{\bf Power Spectra.} {Panel {\sl a)}; Lomb-Scargle power spectrum of the entire $V$ light-curve. Panel {\sl b)}; Lomb-Scargle power spectrum of the HeII radial velocities curve. The dotted line shows the position of the highest peak of the $V$ band periodogram ($P$\,=65.165\,d).}}
\end{figure}

\begin{figure}[H]
\caption{{\bf UVOT multiband photometric light-curve.} Times of photometric maxima used to constrain the superorbital period in 2012 and 2013 are shown with arrows. Black squares: {\it u} ($\lambda_{\mathrm central} $\,=\,3465\AA). Lozenges:  green: {\it uw1} ($\lambda_{\mathrm central} $\,=\,2600\AA), red: {\it um2} ($\lambda_{\mathrm central} $\,=\,2246\AA),  blue: {\it uw2} ($\lambda_{\mathrm central}$\,=\,1928\AA).}
\end{figure}

\begin{figure}[H]
\caption{{\bf Orbital and superorbital periods.} Best fitting orbital \Porb\,=\,63.52\,d) and superorbital \Ps\,=\,2,620\,d $\approx$ 7.2\,yr) solutions accounting for the periodic phase jitter of times of optical/UV photometric maxima. Error bars show 1$\sigma$ statistical errors. Five times of maximum light were extracted from the $V$ band photometry at MJD = 53314.8\,$\pm$\,3, 53636.4\,$\pm$\,3, 53699.4\,$\pm$\,2 (Las Campanas), 55532.8\,$\pm$\,3 and 55788.0\,$\pm$\,2 (ESO-VLT) and four from the UVOT photometry at MJD = 55468.0\,$\pm$\,2 ({\it u}), 56175.0\,$\pm$\,3  ({\it u}), 56243.0\,$\pm$\,2  ({\it u, um2,uw2}) and 56303.0\,$\pm$\,2 ({\it u, um2,uw2}).} 
\end{figure}

\begin{figure}[H]
\caption{{\bf Folded radial velocity curves.} {HeII $\lambda$4686 (panel {\sl a)}) and Balmer absorption (panel {\sl b)}) radial velocity curve folded with the best combination of orbital and superorbital periods (\Porb\,=\,63.52\,d and \Ps\,=\,2620\,d). Phase 0 corresponds to maximum light. Blue: ESO 2009, Black: ESO 2010. Red: ESO 2011. The HeII line displays a clear velocity change with orbital phase. Error bars in all panels show statistical error uncertainties at the 1$\sigma$ level.}}
\end{figure}

\begin{figure}[H]
\caption{{\bf Folded light-curves.} {{\it u}, $V$ and $V$--$I$ light-curves folded with the best combination of orbital and superorbital periods (\Porb\,=\,63.52\,d and \Ps\,=\,2,620\,d). Phase 0 corresponds to the predicted time of maximum optical light. Note the different scales used to plot {\it u} and $V$ light-curves. Panel {\sl a)} {\it u}; black: 2010 X-ray bright state run, red: data acquired during faint X-ray state in the time interval 2012-09-02 to 2013-01-12. Panel {\sl b)} $V$; black: Las Campanas, red: ESO 2010 (X-ray bright), blue: ESO 2011 (X-ray faint). Panel {\sl c)} shows the binned $V$--$I$ light-curve. $V$--$I$ index is plotted on an arbitrary scale. Error bars in all panels show statistical error uncertainties at the 1$\sigma$ level.} }

\end{figure}

\begin{figure}[H]
\caption{{\bf High order Balmer lines.} Normalized mean P13 spectrum outside maximum light ($V$\,$\geq$\,20.3\,mag, shifted up by 0.25 for clarity) compared to that of the B8Ia star $\beta$\,Ori (shifted down by 0.2). The equivalent widths of the high order Balmer lines H8 and H9 are almost identical in both stars. The higher interstellar absorption toward P13 than toward Orion is responsible for the stronger CaII line in the ULX spectrum. Residual Balmer emission already adds to the photospheric H$\delta$ absorption line and the H$\epsilon$ line is blended with the Ca\,II interstellar line. The mean EW of the H8 and H9 Balmer lines are 1.52\,$\pm$\,0.09 and 1.56\,$\pm$\,0.08 for $\beta$\,Ori and P13 respectively (1\,$\sigma$ errors) consistent with no line veiling in P13.}
\end{figure}

\clearpage
\setcounter{page}{1}
\begin{center}
{\Large Supplementary Information} \\
\end{center}

\headertitle{Supplementary Information}
\mainauthor{Motch et al.}

\tableofcontents

\section{Optical spectroscopic observations}

Our spectroscopic monitoring spans several ESO periods from late 2009 until late 2011. 
All ESO-VLT observations were acquired with FORS2 equipped with grism 1200B yielding a resolution of $\approx$ 1,800 at 4000\AA\ for a 1\farcs 0 slit and covering the wavelength range from 3730 to 5190\AA. This setup is well suited for the measurement of small radial velocity variations. Only 5 of the 19 observations granted in P84 could be executed in the nominal P84 period (2009). The programme was pursued in P85 and P86 (2010) and a total of 11 observing blocks were performed. Finally, we obtained additional monitoring in P87 (2011) consisting in 20 observations which were executed in a mostly satisfactory manner. Calibration data were acquired following the standard calibration plan which includes bias, flatfield and arc measurements at the end or at the beginning of the observing night. The same slit width (1\farcs 0) and orientation was used for all observations. This slit position overlaps with a line emitting region close to P13 whose emission lines could be used to monitor the accuracy of the wavelength calibration. We obtained a B filter through-slit image before starting the spectroscopic observations in order to correct velocities for possible off-centre position of the target. The vast majority of the spectra were acquired with a seeing better than 1\farcs 2. 

The standard ESO reduction pipeline does not yield a correct wavelength calibration in the blue part of the 1200B grism spectrum, due to paucity of reference lines in the list of arc calibration lines. We found that the velocity of the high-order Balmer lines ($\lambda\,\leq\,3900$\AA) was shifted by $\approx$ 60\,\kms\ with respect to the rest of the spectrum ($\lambda\,\geq\,4000$\AA). In order to recover an accurate wavelength solution, we used a standard MIDAS procedure to recalibrate the wavelength calibrated 2-d frames produced by the ESO pipeline, taking into account additional arc UV lines for a total of 18 reference arc lines. Wavelengths solutions have rms dispersion of the order of 0.02\AA\ ($\approx$ 1.5\,\kms). One-dimensional spectra were then extracted from the resulting frames using a method optimised for providing the best signal to noise ratio. We automatically rectified all one-dimensional spectra by fitting a spline to continuum points averaged over wavelength intervals free of absorption or emission lines.

We checked the final wavelength calibration by carefully measuring the position of the nebular \Hbet\ emission line present in the local sky background and of the CaII K lines which depending on sky conditions were detected at Galactic velocities or at the velocity of NGC\,7793 in sky spectra. The velocity dispersion of the sky lines was $\approx$ 5\,\kms\ in 2009 and 2010 while it reached a value of $\approx$ 10\,\kms\ in 2011. 

Line velocities were computed using a dedicated cross-correlation tool. The cross-correlation was only computed on specific wavelength windows designed to maximise the signal and exclude lines with possible interstellar origin, such as the Ca II lines for instance. The choice of the 'zero velocity' template spectrum had little impact on the shape of the velocity curves. We tried the direct sum of all observed  spectra, the velocity shifted sum, as well as the observed spectrum of the B8I star $\beta$\,Ori extracted from an early type star atlas\citebraket{\cite{walborn1990}}. {We did not find any suitable digital B9I template. However, the differences between B8I and B9I high-order Balmer lines profiles and depths are small enough to have negligible effects on the quality of the cross-correlation}. Slightly better results were obtained using a TLUSTY BSTAR06\citebraket{\cite{lanz2007}} model atmosphere with \Teff\,=\,15,000\,K, $\log g$=2.0 degraded at the resolution of the grism 1200B observations of P13. \Teff\,=\,15,000\,K is the lowest temperature available in the BSTAR06 collection. Although the model \Teff\ appears slightly higher than that finally derived from the spectral type determination (\Teff\,=\,11,000\,$\pm$\,1,000\,K), the high-order model Balmer lines have equivalent widths fully consistent with those observed from P13. Therefore, this validates the use of the BSTAR06 as reference spectrum for the cross-correlation. 

We computed the cross correlation functions using small spectral intervals around lines of H12, H11, H10, H9, H8, HeI and H$\delta$. H$\epsilon$ overlaps with the Ca II line while H$\gamma$ already shows evidence of very significant re-emission in its profile and was discarded on this basis. We obtained consistent results even when restricting the range of lines entering the computation of the cross-correlation function. We measured velocities of the HeII $\lambda$4686 emission 
line by fitting Gaussian profiles to the broad and dominating component of each spectrum or again, by using the same cross-correlation tool as for absorption lines with the sum (raw or velocity corrected) profile as template. Consistent velocities were obtained using the two methods. We also corrected the measured velocities for the small de-centring of the stellar image in the 1\arcsec\ large slit yielding a rms correction of $\approx$ 12\,\kms\ and eventually moved the velocities to the solar system barycentre.

We estimated errors on radial velocities by using a Monte Carlo procedure, based on the statistics of the original data, taking into account object + sky counts only and applied to the entire processing chain starting from the spectrum normalisation process. 

\section{Optical Photometry}

We monitored the photometric variability of the optical counterpart in the time interval from 2004 to early 2012. About half of our photometric data comes from observations carried out with the Warsaw 1.3m telescope at the Las Campanas Observatory, while the rest consists of data obtained at ESO as part of the acquisition process of the spectroscopic observations (see. 
Extended Data Fig.~\ref{figext:opticaldata}).

\subsection{Las Campanas observations}
The camera consisted of a mosaic 8k\,$\times$\,8k detector providing a plate scale of 0\farcs25\,pixel$^{-1}$ over the 35\arcmin\,$\times$\,35\arcmin\ field of view. A total of 45 $\times$ 900\,s long exposures in the $V$ and $I$ filters were obtained over a time interval of 454 days, from September 2004 until December 2005.
We calibrated the Las Campanas differential magnitudes by fitting the mean photometric level of the light-curve maxima measured in the ESO light-curve (see below). 

\subsection{ESO observations}

Our ESO $V$ photometry is based on the acquisition frames used in 2010 and 2011 to centre the target in the slit of the FORS2 instrument.
In all cases, we used a plate scale of 0\farcs126\,pixel$^{-1}$, the {\em v\_high} filter with an exposure time of 60\,s in 2010 and 120\,s in 2011. The distribution in time of the photometric light-curve is therefore similar to that of the spectroscopic data. Since the standard ESO pipeline does not process acquisition images, we had to retrieve associated calibration files from the ESO archive (individual flats and biases obtained at times as close as possible to those of the observations and acquired with the same instrumental setting). Night averaged flat fields and bias were accumulated using the ESO pipeline {\it esorex} software and standard FORS2 recipes. Final image correction was also performed with the help of FORS2 {\it esorex} tasks. Differential magnitudes were computed with Sextractor\citebraket{\cite{bertin1996}} using 6 non-saturated comparison stars. In most cases, the FWHM seeing was significantly less than the required maximum 1\farcs2 value. The scatter in magnitude difference between bright comparison stars was found to be the smallest ($\approx$ 0.03 mag) when using a 5 pixel diameter aperture (1\farcs26).

Photometric calibration was done using clean HST/ACS drizzled images in the F435W and F555W filters matching part of the VLT images and obtained from the Hubble Legacy Archive. HST images were obtained in 2003 and have an exposure time of 680\,s each (Program ID 9774, PI: Larsen). The magnitudes of the comparison stars were computed using {\em daophot}, with an aperture of 0\farcs2 and correcting for the missing part of the PSF and transformed to the Johnson's $B$ \& $V$ magnitude scale using\citebraket{\cite{sirianni2005}}. The error on the HST/ACS to $V$ band magnitude transformation is about 0.01 mag, while the mean magnitude of the 6 comparison stars was measured with an accuracy of 0.012 mag in $V$ yielding a FORS2 $V$ band zero point accurate to 0.015 mag. 

\section{X-ray observations}

\subsection{{\it Chandra}}

NGC\,7793 was observed by {\it {\it Chandra}}/ACIS-S on 2003 September 7 (ObsID 3954, 49ks, from the public archive), on 2011 August 13 (ObsID 14231, 59 ks, PI R. Soria), and on 2011 December 25 (ObsIDs 13439 and 14378, 58+25 ks, PI R. Soria). We filtered the event files and extracted images and spectra with the standard {\it dmcopy} and {\it specextract} tasks from the {\footnotesize {CIAO}} data analysis system, Version 4.5\citebraket{\cite{fruscione2006}}. We modelled the spectra with {\footnotesize {XSPEC}} Version 12.7.1\citebraket{\cite{arnaud1996}}. 

\subsection{{\it XMM-Newton}}

NGC\,7793 was first observed by {\it {\it XMM-Newton}} on 2012 May 5 (OBSID: 0693760101). We used SAS version 11.0 to reduce these EPIC data. Only 20\,ks of the 50\,ks long exposure was free of strong soft proton contamination. A second observation took place on 2013 November 25 (OBSID: 0693760401). EPIC data were reduced using SAS version 13.0. Total exposure time was 41.1\,ks and 47.1\,ks for the EPIC pn and MOS cameras respectively.  
All {\it XMM-Newton}/EPIC observations were acquired through the medium filter using the full window modes. Calibrated event lists were created using the SAS {\em emproc} and {\em epproc} procedures. SAS tasks {\it arfgen} and {\it rmfgen} computed proper transmission curves and response matrix files at source position. EPIC camera spectra were merged using the {\em epicspeccombine} task. We checked that spectral fits performed on the 3 EPIC cameras simultaneously provided fully consistent results. {\footnotesize {XSPEC}} Version 12.7.1 was used to model EPIC spectra. 

\subsection{{\it Swift}}

We first obtained a series of 17 {\it Swift}/XRT pointings in 2010 totalling 54.5\,ks of effective exposure time and covering a time interval of 73 days in support of our ESO spectroscopic campaign (see Extended Data Fig.\ref{figext:Swift}). Due to scheduling problems at ESO, optical observations could only start near the end of the X-ray monitoring. Data reduction was performed using the on-line {\it Swift}/XRT data product generator\citebraket{\cite{evans2009}} at the UK {\it Swift} Science Data Centre at the University of Leicester\footnote{http://www.swift.ac.uk/user\_objects/}.  
Additional observations were scheduled in 2011, 2012 and 2013 with the aim to monitor the behaviour of the source at some key orbital phases simultaneously with observations at ESO and to trigger pending {\it XMM-Newton} observations. 

\section{UV observations}

We downloaded the {\it Swift}/UVOT observations from the HEASARC data archives. We used standard {\small {STARLINK}} software {\it Gaia} to display the images and perform aperture photometry on the source. We extracted the source counts from a circle of radius 5 arcsec, and the local background from a suitable annulus (9 to 12 arcsec); we used other nearby, brighter sources to estimate the aperture correction. We converted from count rates to fluxes and Vega magnitudes using the zeropoints tabled in \citebraketlow{\cite{poole2008}}. Over the time interval 2010-2013 we collected 6 images in the {\it uw2} ($\lambda_{\mathrm central}$\,=\,1928\AA), 7 in the {\it um2} ($\lambda_{\mathrm central} $\,=\,2246\AA), 3 in the {\it uw1} ($\lambda_{\mathrm central} $\,=\,2600\AA) and 23 in the {\it u} band ($\lambda_{\mathrm central} $\,=\,3465\AA).  

\section{The ultraluminous state in P13}\label{ulxstate}

The November 2013 {\it XMM-Newton}/EPIC spectrum cannot be fitted by a simple powerlaw neither entirely nor at E\,$\geq$\,1.5\,keV (null hypothesis probabilities of 3.4\,$\times$\,10$^{-5}$ and 8.1\,$\times$\,10$^{-4}$ respectively). {An F-test shows that a broken power-law brings a very significant improvement above 1.5\,keV (see Extended Data Table 1), with a null probability of $1.5\times 10^{-19}$.} However, a simple broken power-law cannot properly fit the entire $0.3$--$12$\,keV energy distribution (null hypothesis probability of 3.5\,$\times$\,10$^{-3}$) due to the presence of an additional soft component.  A disk blackbody added to the broken power law can adequately represent the entire energy distribution.

While these descriptions provide good fits to the data, they are phenomenological rather than physical models; moreover, in such models, the power-law is unphysically extended at low energies, below the energy of the seed thermal component. That is why physical Comptonization models should be used instead. A {\it diskbb} + {\it comptt} model provides an excellent fit to the {\it XMM-Newton} spectrum, and a similarly good fit is obtained with the {\em diskir} model\citebraketd{\cite{gierlinski2008}}{\cite{gierlinski2009}} (see Extended Data Fig.~\ref{figext:ChandraXMMSpec}). The main difference between the two Comptonization models is that {\it comptt} assumes seed photons with a Wien distribution, and the direct disk-blackbody emission (typical of X-ray binary spectra) must be included as a separate component; instead, in {\it diskir} a multi-colour disk-blackbody is self-consistently used for the input spectrum. The quality of the {\it XMM-Newton} data was not high enough to independently fit the seed photon temperature $kT_0$ in {\it comptt} and the peak colour temperature $kT_{\mathrm in}$ in {\it diskbb}. Following standard practice we assumed $kT_0$ = $kT_{\mathrm in}$. In super critical conditions, the peak temperature of the thermal component (here $kT_{\mathrm in} \approx 0.2$--$0.3$\,keV) is interpreted as the disk temperature at the spherization radius, inside which a geometrically-thin standard disk is no longer viable and the outflowing scattering region starts\citebraket{\cite{poutanen2007}}. The normalization constant in {\it diskir} provides an approximate estimate of the inner disk radius: $R_{\mathrm in} \approx 1.2 \left(N_{\rm dbb}\right)^{1/2} \, \left(d/10{\rm kpc}\right)\, \left(\cos i\right)^{-1/2} \sim 2,000$ km\citebraket{\cite{kubota1998}}. 
We interpret this as the inner radius of the standard disk being located  far from the innermost stable circular orbit. Large ($> 1,000$\,km) characteristic radii of the soft thermal component are one of the defining 
properties of most ULXs, and they were initially interpreted as evidence of intermediate-mass black holes\citebraketd{\cite{miller2003}}{\cite{miller2004}}

A single power law fit to the 2003 {\it Chandra} data yields a photon index $\Gamma$ $\approx$ 1.19\,$\pm$\,0.05 (null hypothesis probability of 0.05). {However, a phenomenological broken power-law model, or a more physical Comptonized model ({\it diskbb} + {\it comptt} or {\it diskir}; see Extended Data Fig.~\ref{figext:ChandraXMMSpec}) considerably improve the fit quality and yield parameters similar to those obtained by fitting the {\it XMM-Newton} spectrum. An F-test shows that {above 1.5\,keV} the medium energy break is statistically significant at the 99\% level. Because of the lower signal-to-noise ratio of the {\it Chandra} data at energies $\lesssim 0.5$ keV, the relative contribution of the direct {\it diskbb} component with respect to the seed thermal component inside {\it comptt} is not well constrained; in fact, the best-fitting parameters are obtained for a pure {\it comptt} model, with {\it diskbb} normalization going to zero, but a {\it diskbb} normalization similar to that found for the {\it XMM-Newton} spectrum is statistically equivalent.
Using the {\it diskbb} + {\it comptt} fit to the {\it Chandra} 2003 data yields a $0.3$--$10$\,keV flux (corrected for interstellar absorption) of 2.11\,$\times$\,10$^{-12}$\,\ergscm\ corresponding to \Lx\,=\,3.5\,$\pm$\,0.1\,$\times$\,10$^{39}$\,\ergs\ ($d$\,=\,3.7\,$\pm$\,0.1\,\,Mpc\citebraket{\cite{radburn2011}}). } Integrated over the same energy range, the {\it diskir} model yields \Lx\,=\,3.4\,$\times$\,10$^{39}$\,\ergs. In order to compute X-ray heating effects we extrapolated the {\it diskbb} + {\it comptt} model up to 20\,keV and obtained \Lx\,=\,4.2\,$\times$\,10$^{39}$\,\ergs. P13 was somewhat fainter in 2013 with \Lx\,=\,2.0\,$\pm$\,0.1\,\,$\times$\,10$^{39}$\,\ergs ($0.3$--$10$\,keV). 

The overall shape of the {\it Chandra} and {\it XMM-Newton} spectra appear quite similar, except for a slightly steeper slope (higher photon index) in the {\it XMM-Newton} spectrum (see Extended Data Fig.~\ref{figext:ChandraXMMSpec}), which may be a function of the lower X-ray luminosity. For the Comptonization component, using the {\it diskir} model, the electronic temperature was $kT_e = 1.64^{+0.12}_{-0.11}$\,keV for {\it Chandra} and $kT_e = 1.80^{+0.12}_{-0.16}$\,keV for {\it XMM-Newton}; using the {\it diskbb} + {\it comptt} model, we found $kT_e = 2.11^{+0.37}_{-0.23}$\,keV for {\it Chandra} and $kT_e = 1.67^{+0.20}_{-0.12}$\,keV for {\it XMM-Newton}; the characteristic optical depth is $\tau \approx 10$ in both spectra. These temperatures are much lower than in the sub-Eddington low/hard state of black hole X-ray binaries, and the optical depth is much higher\citebraket{\cite{remillard2006}}. 

This medium energy break is thought to be due to strong Comptonization of low energy photons in a dense, { {relatively cool}} corona\citebraket{\cite{gladstone2009}} or in an expanding wind with electronic temperatures of 1-3\,keV\citebraket{\cite{roberts2005}}. Alternatively, the break may be the peak of non-standard disc emission, distorted by overheating, spin, photon trapping or turbulence, for example emission from a slim disk\citebraket{\cite{watarai2001}}. 
We conclude that the bright state X-ray spectrum of P13 definitely shows the two hallmarks (soft thermal multicolour excess  with a large characteristic size, and spectral downturn at a few keV) of the canonical ultraluminous X-ray state visible in most ULX with available high signal to noise spectra\citebraket{\cite{gladstone2009}}. 

\section{X-ray bright and faint states}

P13 could not be detected by the {\it Swift}/XRT in any of the 24 individual observations obtained from 2011 August 26 until 2013 June 3. The ($0.3$--$10$\,keV) X-ray luminosity observed by {\it Chandra} in 2011 was \Lx\,$\approx$  5\,$\pm$\,1\,$\times$\,10$^{37}$\,\ergs\, consistent with being the same on 2011 August 13 and December 25. Similarly, on 2012 May 14, {\it XMM-Newton} detected P13 as a faint X-ray source with \Lx\,($0.3$--$10$\,keV)\,=\,4-5\,$\times$\,10$^{37}$\,\ergs. Finally, summing the 19 {\it Swift}/XRT snapshot exposures accumulated between 2012 July 20 and  2013 January 18 for a total of 45.5\,ks allows us to detect P13 with a count rate of 3.1\,$\pm$\,0.7\,$\times$\,10$^{-4}$\,\cnts, corresponding to \Lx\,$\approx$\,5\,$\pm$\,1\,$\times$\,10$^{37}$\,\ergs. The faint state X-ray luminosities are about two orders of magnitude lower than those observed in 2003 and late 2013. P13 was recovered by {\it Swift}/XRT on 2013 November 19 with a count rate of $\approx$ 0.022\,$\pm$\,0.003\,\cnts. 

\section{Determination of the orbital period}

A Lomb-Scargle\citebraket{\cite{lomb1976}} power spectrum of the $V$ light-curve is plotted in Extended Data Fig.~\ref{figext:power_Vheii} panel {\it a)}. The two highest peaks correspond to a beating phenomenon between the Las Campanas and ESO data and cannot be prioritised on a statistical basis. We estimate the probability that such high peaks are due to red noise fluctuations at less than 0.3\% when considering all possible independent frequencies searched from 0 to 0.1 cycles per day. Local red noise power was estimated by removing the main modulation and its first harmonics and averaging residual power at periods longer than 16\,d.  The longer period at $P$\,=\,65.165\,d exhibits slightly more power than the second candidate period $P$\,=\,63.340\,d. Treated independently, the Las Campanas and the ESO data sets yield as best periods $P$\,=\,64.75\,$\pm$\,0.75\,d and $P$\,=\,64.0\,$\pm$\,1.4d respectively. 

The Las Campanas $V$--$I$ index time series obtained in 2004 and 2005 also displays significant power at a period of 65.6\,$\pm$\,1.8d, formally consistent with the two 65.165\,d and the 63.340\,d possible periods derived from the analysis of the entire Las Campanas and ESO $V$ time series. 

Importantly, the He\,II $\lambda$4686 velocity also shows a clear signal in this range of orbital periods. The Lomb-Scargle power spectrum shown in Extended Data Fig.~\ref{figext:power_Vheii} panel {\it b)} indeed exhibits a peak at $P$\,=\, 65.5\,$\pm$\,1.4\,d, similar to that provided by the $V$--$I$ time series, thus favouring the long photometric period but still being consistent with the shortest solution. The probability that the He\,II $\lambda$4686 velocity randomly exhibits an equally high modulation at the same frequency as the photometry is $\approx$\,3\,$\times$\,10$^{-4}$. 

At any rate, it is clear that the phase of maximum light in the $V$ light-curve shows considerable variability on a time scale of several months. None of the candidate periods yields a smooth folded light-curve. In all cases low data points are found either on the ascending branch to maximum light or on the descending branch.

Although the time coverage is not as good as for the {\it u} band, large intensity changes are also observed in the shorter wavelength filters. By combining all UVOT filter data and $V$ band time series, a total of 9 times of maximum light could be unambiguously identified over a time interval from November 2004 until January 2013. Extended Data Fig.~\ref{figext:UVOTmaxima} shows the position of photometric maxima determined from multiband UVOT observations in 2012 and 2013. No constant periodicity can fit all maxima, thus confirming the intrinsic phase variability suspected based on $V$ band data alone. Maxima occur with time lags of up to $\pm$\,6\,days with respect to a constant period. Introducing a period derivative provides an acceptable fit to the series of times of maxima. However, the resulting value of $P_{0}$/$\dot{P}$ of $\approx$ 190\,yr is extremely small for any evolutionary time scale. Alternatively, we assumed that the time of photometric maxima were shifted regularly back and forth following a superorbital clock. We tested orbital periods in the range of 63 to 66 days and superorbital periods in the range of 200 to 6,000 days. We selected the combination of orbital and superorbital periods yielding the minimum $\chi^{2}$ deviation to the observed times of maximum light. In order to improve on the accuracy of the times of maxima in the $V$ band we cross-correlated the binned folded light-curve corrected for phase jitter with that observed around individual maxima.
Three iterations were enough to converge toward a best stable solution with an orbital period of 63.52\,d and a superorbital period of 2,620\,d (see Extended Data Fig.~\ref{figext:PeriodFitResiduals}). We underline that this combination of orbital and superorbital periods is only the best one among many others. At the 68\% confidence level, the vast majority of acceptable superorbital periods are in the range of $\approx$ 1,800\,d to $\approx$ 3,200\,d. Some acceptable combinations of orbital and superorbital periods also exist at $\approx$ 1,300\,d and 4,500\,d. Corresponding orbital periods are 63.44\,$\pm$\,0.24\,d and 65.16\,$\pm$\,0.16\,d.  
In all cases, applying this correction allows us to remove almost all out of phase photometric measurements and accumulate a much cleaner light-curve than was possible assuming no superorbital periodicity. Since all statistically acceptable combinations of orbital and superorbital periods yield similar time dependent phase shifts, all resulting folded light-curves look very similar. For the sake of the light-curve fitting, we will keep the one obtained for the best solution quoted above.

\section{Optical behaviour}

Contemporaneous X-ray observations reveal that while the X-ray source was bright in 2010, its luminosity had dropped by two orders of magnitude in 2011. In spite of this large change in observed X-ray luminosity, the equivalent width of the He\,II $\lambda$4686 emission line and the amplitude of the $V$ light-curve, both believed to be due to X-ray heating, remained unchanged (see Extended Data Fig.~\ref{figext:opticaldata}). However, the change in X-ray state had a strong impact on both absorption and emission lines velocity patterns. Whereas in 2010, the Balmer absorption lines displayed a velocity change in roughly opposite direction as to the He\,II line, its variability pattern became apparently erratic and unrelated to orbital phase in 2011 (see Extended Data Fig.~\ref{figext:foldedHeII}). Similarly, the 2011 He\,II radial velocity curve displayed a maximum receding velocity at a later phase with respect to optical maximum light than in 2010, whatever is the orbital period considered, 63.340\,d or 65.165\,d. 

Extended Data Fig.~\ref{figext:foldedLC} shows the {\it Swift}/UVOT {\it u}, $V$ and $V$--$I$ light-curves folded with an orbital period of 63.52\,d with time jitters corrected for a superorbital period of 2620\,d. Total amplitudes of $\approx$ 0.5 mag, $\approx$ 1.0 mag $\approx$ 1.3 mag and $\approx$ 1.5 mag are observed in the $V$, {\it u}, {\it um2} and {\it uw2} filters respectively. The instrumental $V$--$I$ light-curve also shows that the star appears bluer at optical maximum. The increasing photometric amplitude with decreasing wavelength underlines that X-ray heating effects are pronounced and responsible for the orbital modulation. 

\section{Disk contribution to the optical light}

In a first step, we used the results of the {\em diskir} model fit to the 2003 {\it Chandra} data to estimate the possible disk contribution to the optical flux. This model implements a scenario in which the outer disk is heavily irradiated by a fraction of the X-ray luminosity, and therefore hotter and more luminous than a standard viscous-heated disk. We assumed \Av\,=\,0.16, $e$ = 0.27, $i$ = 50\,\degr, a B9Ia mass of 23\,\Msol\ and a maximum disc radius of 90\% of the equivalent Roche lobe radius at periastron, implying $\log r_{\mathrm{out}}$\,=\,4.56. The most sensitive quantity is $f_{\mathrm out}$, the fraction of the bolometric flux thermalized in the outer disk. Its highest recorded value is $\approx$ 5\% in ULX  NGC\,5408\,X-1\citebraket{\cite{grise2012}}. Assuming $f_{\mathrm out}$ = 5\% we find that the disk contribution to the optical luminosity remains about one mag below the stellar contribution (i.e. $V$\,=\,21.5\,mag). We note that model parameters $\log r_{\mathrm{out}}$ and $f_{\mathrm{out}}$ have no effects on the X-ray spectral shape and only impact the optical energy distribution. 

In a second step, we estimated the possible contribution of the accretion disk to the $V$ band by accumulating all spectra obtained outside maximum light ($V$\,$\geq$\,20.3\,mag). At these phases, emission from the heated parts of the star is minimum so that the only possible additional $V$ band light may come from the accretion disk. We then compared the Balmer line equivalent widths observed in P13 with those seen in the template B8Ia star $\beta$\,Ori\citebraket{\cite{walborn1990}} (see Extended Data Fig.~\ref{figext:EWcomparison}). The effective temperature corresponding to B8Ia is only 1,000\,K hotter than that of a B9Ia star. Comparing the mean EW of the H8 and H9 Balmer lines shows no evidence of veiling. At the 3\,$\sigma$ confidence level, disk emission in the B band could at most be $\approx$ 35\% of the stellar continuum, about half of the maximum predicted by the {\em diskir} model. In order to be on the safe side, we will retain a maximum light contribution for the accretion disk corresponding to a $V$ magnitude of 21.5.

\section{Light-curves fitting}

All light-curves were folded assuming \Porb\,=\,63.52\,d and arrival times were corrected with a superorbital period of 2620\,d.  
The $V$ light-curve was folded into 20 phase bins. Although a periodic shift of times of photometric maxima has considerably reduced light-curve scatter, noticeable cycle to cycle variations remains.  With the set of orbital and superorbital periods considered here, variability is particularly strong in the rising phase of the $V$ flux to maximum (see Extended Data Fig.~\ref{figext:foldedLC}) and remains at a lower level all along the rest of the light-curve. { This scatter may be either due to some remaining red noise (e.g. caused by the variable X-ray source) or may reflect the complexity of the changes of the shape of the light-curve with superorbital phase.} 
In order to account for this effect, we quadratically added a phase dependent intrinsic scatter to that derived from bin averaged measurements. This phase dependent intrinsic scatter was computed by comparing the variance expected from measurement errors with that in phase bins moving around the central phase value. Owing to the scarcer coverage of the UVOT data (only 23 measurements available in the {\it u} band), we directly fitted the individual folded data measurements shown in the upper panel of Extended Data Fig.~\ref{figext:foldedLC}. 

X-ray emission was assumed to be isotropic, i.e. emitted from a small spherical central region rather than from an accretion disk. Stellar photospheric emission was assumed to be blackbody like with a linear limb darkening law of coefficient 0.635 and assuming a gravity darkening exponent of 0.25 suitable for stars with radiative envelopes\citebraket{\cite{orosz2000}}. This was done because the grid of available stellar atmosphere models in ELC did not cover the range of low gravity and \Teff\ needed in the case of P13. We computed the effective UVOT {\it u} wavelength by folding the transmission curve with the theoretical flux spectrum used as velocity reference (\Teff\,=\,15,000\,K and $\log g$\,=\,1.75). The resulting wavelength, $\lambda_{eff}$\,=\,3470\,\AA\ is comparable to the central filter pass band wavelength. 

X-ray heating effects on the supergiant star were computed assuming total absorption of the high energy photons in the stellar photosphere as expected for E\,$\leq$\,30\,keV X-rays deposited at close to normal incidence\citebraket{\cite{george1991}}. Although the shape of the model light-curve hardly depends on the rotation rate of the mass-donor star, the Roche lobe radius of the B9Ia star and therefore its optical luminosity will vary significantly depending on whether it is synchronised with orbital rotation or not (see below section \ref{fbhm}). We tested two extreme stellar rotation rates, first a non rotating star and second a star synchronised with orbital motion at periastron.  

$V$ and {\it u} light-curves were simultaneously fitted for a grid of 39 \Mx/\Mopt mass ratios ranging from 0.1 to 10, 19 systems inclination angles from 0 to 90\,\degr\ with a step of 5\,\degr, two primary masses and four X-ray luminosities. The remaining orbital parameters, eccentricity, periastron angle and phase of periastron were then fitted using the "grid search" routine\citebraket{\cite{bevington1969}}. In order to ensure best convergence, we used as starting parameters results from the last step best fit. Since the contribution of the disk to optical light is expected to be small we assumed a dark disk whose main effect is to cast X-ray shadows on the mass donor star. We fitted the disk $\beta$ half-opening angle while fixing the outer radius at a value of 70\% of the Roche lobe radius at periastron, a value adapted to the case of eccentric orbits\citebraket{\cite{artymowicz1994}}. 

\section{A precessing accretion disk ?}

The persistence of conspicuous X-ray heating effects during X-ray faint states strongly suggests the presence of a partly occulting precessing disk similar to those observed in the well documented X-ray binaries Her X-1\citebraket{\cite{giacconi1973}} and LMC X-4\citebraket{\cite{lang1981}}. We thus expect the shape of the optical light-curve to evolve with precession phase as a result of a changing disk rim height and azimuthal position\citebraketd{\cite{deeter1976}}{\cite{ilovaisky1984}}.  In particular Her X-1 exhibits periodic maximum light phase shifts with the superorbital 35\,d period\citebraket{\cite{deeter1976}}, a situation comparable to that apparently seen in P13. 

The observed ratio between orbital and superorbital periods is $\approx$ 10\,-\,20 for established precessing disks\citebraket{\cite{sood2007}}. For P13, most of the possible superorbital periods are in the range of $\approx$ 1,800\,d to $\approx$ 3,200\,d implying superorbital to orbital period ratios in the range of 28 to 50. These period ratios are slightly higher than those of established precessing disks, but not inconsistent with period ratios seen in the black hole binaries 1E 1740.72942 and GRS 1758-258\citebraket{\cite{smith2002}}. 

Our X-ray monitoring is too sparse to efficiently constrain the duration of the faint state and could have missed a short bright state such as observed in HerX-1\citebraket{\cite{scott1999}}.  The time interval between the first and last observed X-ray faint states is 660\,d while X-ray bright states observations define a maximum faint state duration of 1,118\,d.  A 7.2\,yr cycle could easily account for the $\approx$ 1.8 $-$ 3.0\,yr long persistent faint X-ray state. It is tempting to relate the apparent X-ray on/off cycle and the optical maximum light cycle to the same underlying clock, namely a precessing disk. 

Interpreting X-ray faint states as due to shielding of the X-ray source by a precessing disk implies rather high inclinations. Her X-1\citebraketd{\cite{cheng1995}}{\cite{reynolds1997}} and LMC X-4\citebraket{\cite{vandermer2007}}, the only well established X-ray binaries exhibiting eclipses and superorbital X-ray on/off cycles are seen at inclinations of 84\degr\,$\pm$\,4\degr\ and 68\degr\,$\pm$\,4\degr\ respectively.  
{ The problem of whether the factor $\sim$ 100 flux attenuation in the faint state is the signature of a relatively small Comptonized region or reflects scattering in a dense outflowing wind will be addressed in a follow-up paper. }

\section{Possible eclipses in the 2010 {\it Swift} observation}

On MJD 55492.9 {\it Swift}/XRT detected a very faint X-ray flux of 0.0012\,$\pm$\,0.0007\,\cnts\ consistent with zero flux (see Extended Data Fig.~\ref{figext:Swift}). This event could be the signature of an X-ray eclipse. Unfortunately, {\it Swift} could not be scheduled to observe P13 one orbital cycle before. Nevertheless, the {\it Swift}/XRT measurements constrain the possible X-ray eclipse duration to be shorter than about 7\,days. 

\section{Constraining the mass of the black hole}\label{fbhm}

Our upper limit on the black hole mass in P13 results from three different and complementary sets of constraints. First, we rejected all combinations of inclination, mass ratio, X-ray luminosity, B9Ia stellar mass and effective temperature for which no satisfactory fit of the $V$ and UVOT {\it u} bands light-curves could be reached at the 99.7\% confidence level (equivalent to $\pm$\,3\,$\sigma$ for a Gaussian distribution). As seen on Fig.~3 of the main article, this condition excludes part of the {\it i / \Mx} plane but still allows for quite massive black holes. Increasing the mass of the black hole enlarges the orbital separation, and therefore decreases X-ray heating effects, and yields smaller light-curve amplitudes than observed. This effect can be compensated for by $\it i)$ rising the orbital eccentricity and therefore moving the X-ray source closer to the companion star at periastron $\it ii)$ assuming higher X-ray luminosities and $\it iii)$ using lower effective photospheric temperature. 

The second and most stringent constraint arises from the very high observed optical luminosity of the B9Ia star. This constraint considerably shrinks the parameter space allowed by fit quality criteria only. Such a large star has to fit into its Roche lobe at periastron. For a fixed orbital period and stellar mass, the size of the Roche lobe of the mass donor star relatively slowly decreases with increasing black hole masses. At some stage, the smallest possible radius of the star, taking into account measurement uncertainties, maximum contribution from the accretion disk and residual X-ray heating, becomes larger than the radius of the Roche lobe. Eccentricities in the range of $\approx$ 0.3 to 0.4 are needed to account for the narrow maximum of the optical and UV light-curves. The high orbital eccentricities also considerably reduce the Roche lobe at periastron as compared to the circular case. Non-synchronicity can further decrease or increase the size of the Roche lobe. Whereas the volume-equivalent Roche lobe radius of a star synchronised with orbital angular velocity at periastron is close to that computed for point-like masses\citebraket{\cite{eggleton1983}}, stars rotating significantly slower than orbital may have volume-equivalent Roche lobe radii up to $\approx$ 10\% larger\citebraket{\cite{sepinsky2007}}. Determining the synchronisation state of mass-donor stars in X-ray binaries is observationally challenging. In the few cases for which it could be measured, the ratio of stellar to orbital angular velocity is in the range of $\approx$ 1.0 to 0.67\citebraket{\cite{Koenigsberger2012}}. However, the rotation velocity of the mass-donor star is expected to strongly depend on whether former mass transfer occured. Orbital solutions computed assuming synchronisation at periastron predict maximum projected rotational velocities of $\approx$ 120\,\kms\ and would not have been detected in our Balmer absorption line profiles. We then retained orbital solutions predicting an intrinsic stellar flux consistent with that observed at minimum $V$ light ($V$\,=\,20.50\,$\pm$\,0.03\,mag). We assumed a distance of 3.7\,$\pm$\,0.1\,Mpc\citebraket{\cite{radburn2011}}, \Av\ in the range of 0.07 (Galactic extinction\citebraket{\cite{schlegel1998}} in the direction of NGC\,7793) to 0.25\citebraket{\cite{piet2010}} and a maximum disk magnitude of $V$\,=\,21.5. This yielded \Mv\ in the range of -6.77 to -7.68. Finally, for each solution, we computed the amount of $V$ light due to residual X-ray heating near light-curve minimum (phases 0.4\,-\,0.6) by switching off the X-ray source keeping all other parameters constant in the ELC modelling. Any additional light arising from X-ray heating effects account for less than 0.1 mag at optical minimum.

The model bolometric magnitude was then given by \Mbol\,=\,42.36 $-$ 5$\times$$\log(R_{\mathrm RL}/$\Rsol) $-$ 10$\times$$\log$(\Teff) and forced to be within the allowed range derived from \Mv\ values and \Teff\ dependent bolometric corrections\citebraket{\cite{flower1996}} (BC = -0.26, -0.49 and -0.71 for \Teff\,= 10,000, 11,000 and 12,000\,K, respectively). 

The last constraint arises from the absence of X-ray eclipses longer than $\approx$\,7\,d during the long 2010 {\it Swift} monitoring. This excludes systems with inclinations higher than $\approx$ 80\degr. These inclinations are still compatible with those of X-ray binaries displaying on-off X-ray states due to a precessing disk. Interestingly, the possible eclipse found at MJD 55492.9 in the 2010 {\it Swift}/XRT monitoring occurs close to the predicted time of inferior conjunction near orbital phase 0.5. 

Hints of a coherent orbital signal were present in the 2010 Balmer absorption radial velocities and we had hoped to establish a well-behaved radial velocity curve for P13. However, rather unexpectedly, the 2011 observations did not recover the 2010 variability pattern. 
Therefore, some velocity noise of astrophysical origin (possibly due to circumstellar and circumbinary material) may hide the underlying orbital variation. The temperature of the most X-ray heated parts of the  B9I star facing the black hole may rise by up to $\approx$ 20,000K at periastron. Consequently, the high order Balmer absorption lines emitted by the X-ray heated parts of the star will have a lower EW than those emitted from the unheated hemisphere, so that the observed Balmer radial velocity may mostly reflect that of the part of the star farthest from the black hole and therefore be an upper limit to that of the true centre of mass. 

Close inspection of the Balmer high-order line profiles do show some evidence for re-emission on occasion, albeit at a level which is unlikely to significantly change the mean velocity. Changes of similar amplitude and shapes have been observed in the radial velocity curve of Her X-1 depending on super-orbital phase\citebraket{\cite{hutchings1985}} but to the best of our knowledge, are not very well documented. In addition, if a precessing disk explains superorbital cycles, complex illumination patterns may be projected on the mass donor star. Unfortunately, the ELC code is not yet able to handle titled disk shadows nor can it handle the variation of the equivalent width of the absorption lines with photospheric temperature. 

The B9Ia star radial velocity curve was computed in the ELC code by summing contributions from individual tiles weighted by their emitted B flux so as to match the wavelength range of the high-order Balmer absorption lines used in the observations. In all cases, this correction remains small and very similar results are obtained when using mass centres velocities instead. Velocities are therefore those of the light barycentre but are not corrected for changes in equivalent widths with temperature. 
All orbital solutions acceptable on the basis of their large enough optical luminosities predict Balmer radial velocity amplitudes lower than the observed upper limit of \maxvel\,\kms\ (see Table 1). The mean full amplitude of the acceptable model velocities of the mass-donor star is $\approx$ 50\,\kms . 
Most solutions predict a periastron angle close to 90\degr, i.e. a situation in which the X-ray heated stellar hemisphere points toward earth at periastron. Such a geometry obviously provides the highest photometric amplitude. However, the model radial velocity variation of the He\,II line seems late by $\approx$\,0.3 in phase with respect to the observed radial velocity curve suggesting that the He\,II line does not accurately trace the velocity of the accreting black hole (see Fig.~2). In addition, the Balmer absorption radial velocity curve cannot be fitted by any of the ELC models.

The upper black hole mass limits are reached for a B9Ia star mass of 23\,\Msol\ and for higher X-ray luminosities (Log(\Lx)\,=\,39.775 - 39.925).  In the non rotating case, maximum black hole masses are \Mx\,$\approx$\,6.3\,\Msol\ and \Mx\,$\approx$\,15.0\,\Msol\ for \Mopt\,=\,18\,\Msol\ and \Mopt\,=\,23\,\Msol\ respectively. In the synchronized case, there is no solution with \Mopt\,=\,18\,\Msol\ and the maximum black hole masses is \Mx\,$\approx$\,7\,\Msol. Acceptable solutions exist for the whole range of X-ray luminosities tested. The higher the illuminating X-ray luminosity, the larger the possible separation between the X-ray source and the mass-donor star, and the higher the mass of the black hole. Relatively higher black hole masses are also favoured by high disk optical luminosities which allow for less luminous and smaller B9Ia stars. Higher stellar effective temperatures allow for smaller Roche lobes and higher black hole masses. Therefore, if the intrinsic (non X-ray illuminated) stellar \Teff\ were lower than assumed here, more stringent mass constraints would be derived. 

\clearpage
\centerline{Additional References (see main text for 1-30)}

\clearpage
\centerline{\bf \large Figures}
\setcounter{figure}{0}
\begin{figure}[h]
\centerline{\psfig{file=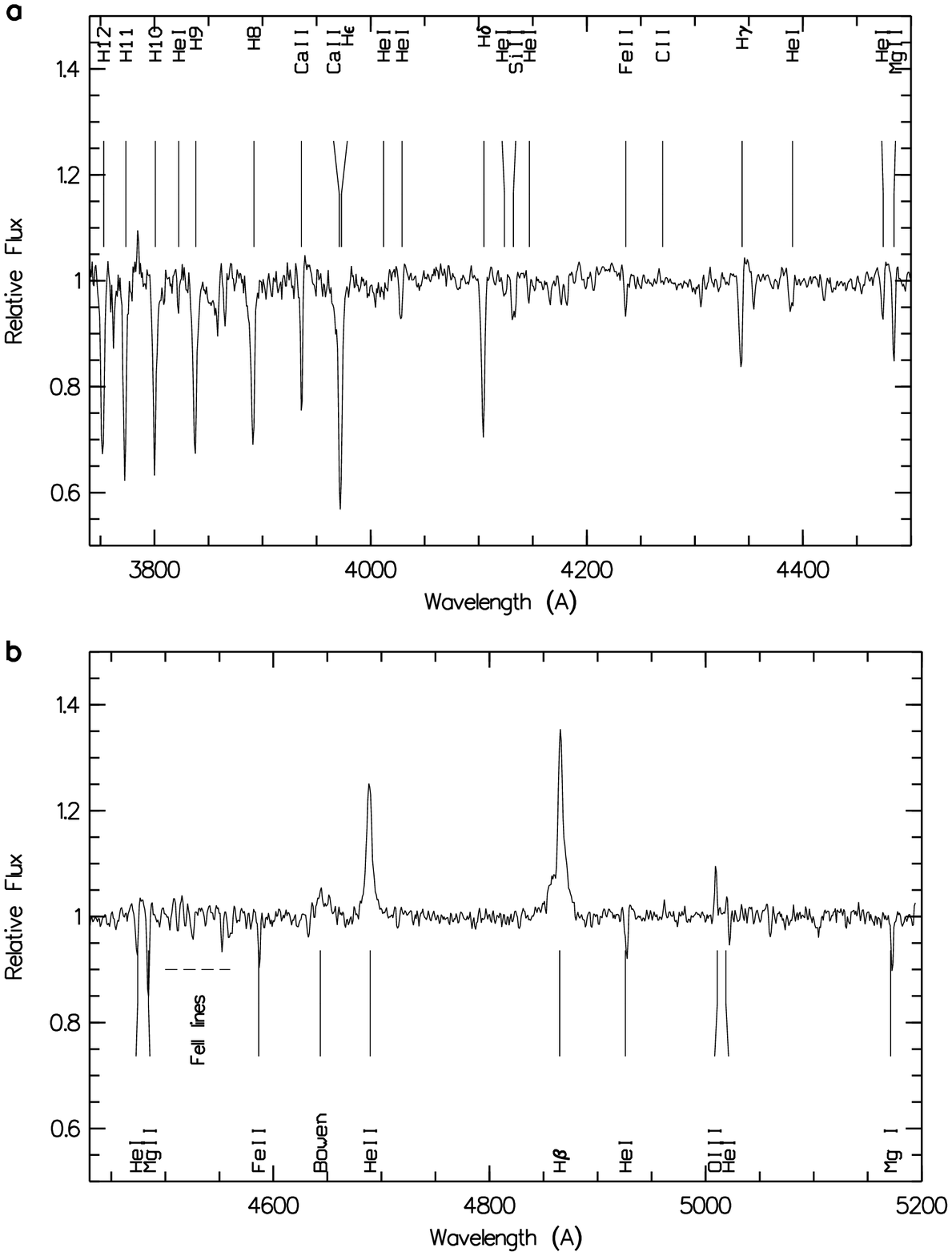,width=14cm,angle=0,bbllx=30pt,bblly=100pt,bburx=550pt,bbury=775pt,clip=true}} 
\addtocounter{figure}{1}
\hbox{\bf Figure \arabic{figure}}
\label{figmain:spec8687}
\end{figure}

\clearpage
\begin{figure}[h]
\noindent \hspace{-1cm}
\centerline{\psfig{file=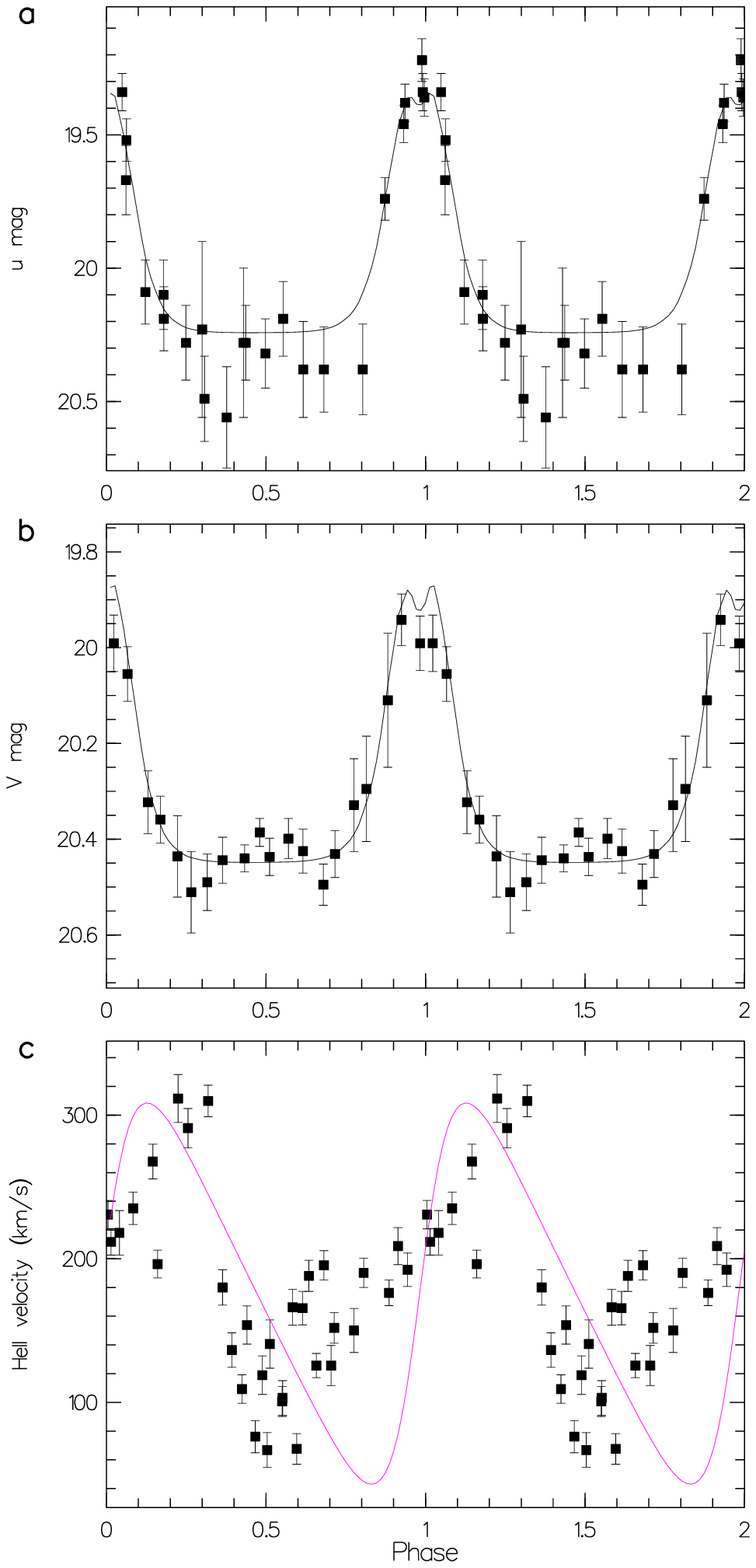,width=17cm,angle=0,bbllx=30pt,bblly=100pt,bburx=550pt,bbury=775pt,clip=true}} 
\addtocounter{figure}{1}
\hbox{\bf Figure \arabic{figure}}
\label{figmain:bestfit}
\end{figure}

\clearpage

\begin{figure}[h]
\centerline{\psfig{file=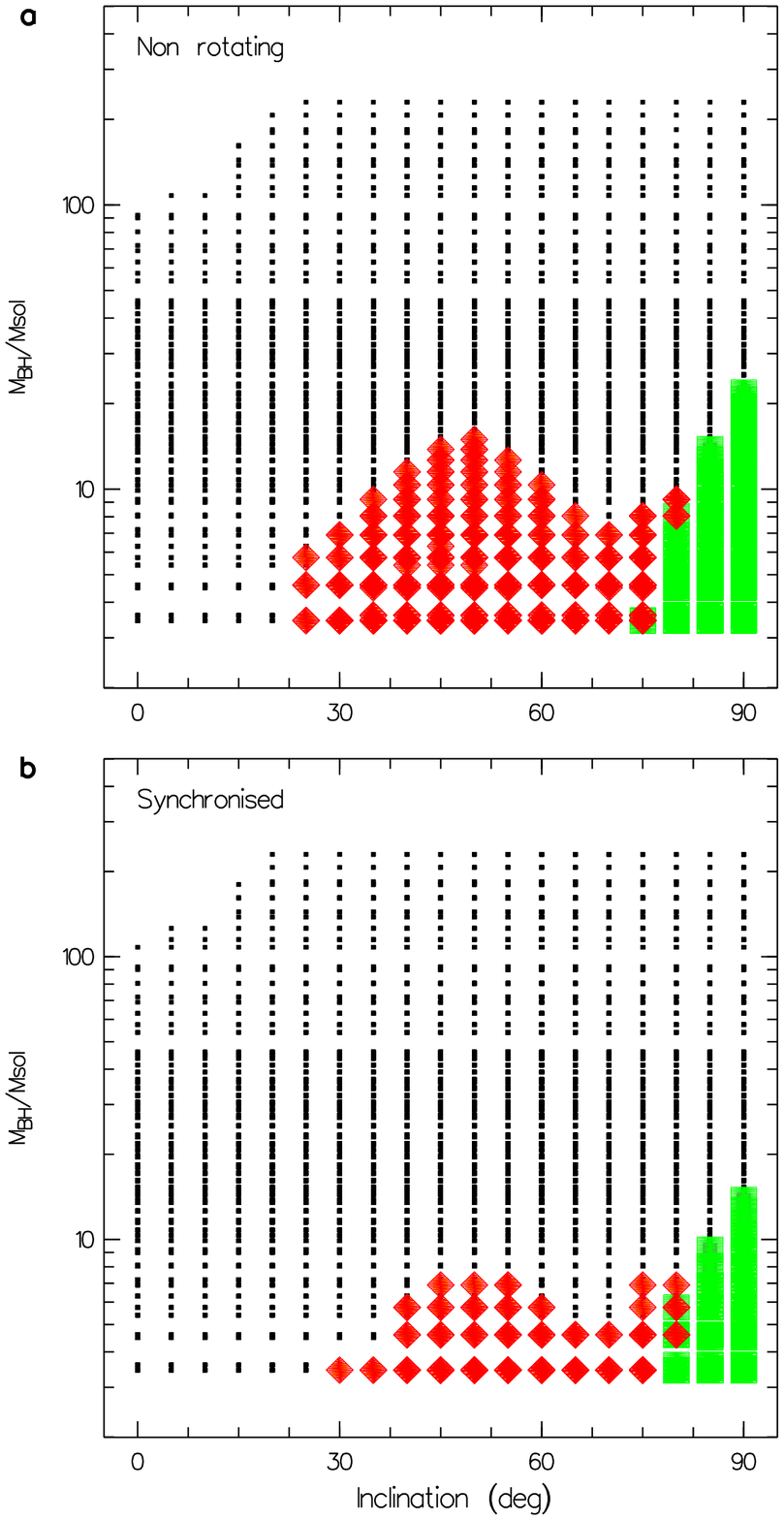,width=17cm,angle=0,bbllx=30pt,bblly=100pt,bburx=550pt,bbury=775pt,clip=true}} 
\label{figmain:AllowedArea}
\addtocounter{figure}{1}
\hbox{\bf Figure \arabic{figure}}
\end{figure}

\clearpage
\centerline{\bf \Large Extended Data}
\setcounter{figure}{0}
\setcounter{table}{0}

\noindent {\bf Extended Data Table 1.}
\vskip 1cm
\psfig{file={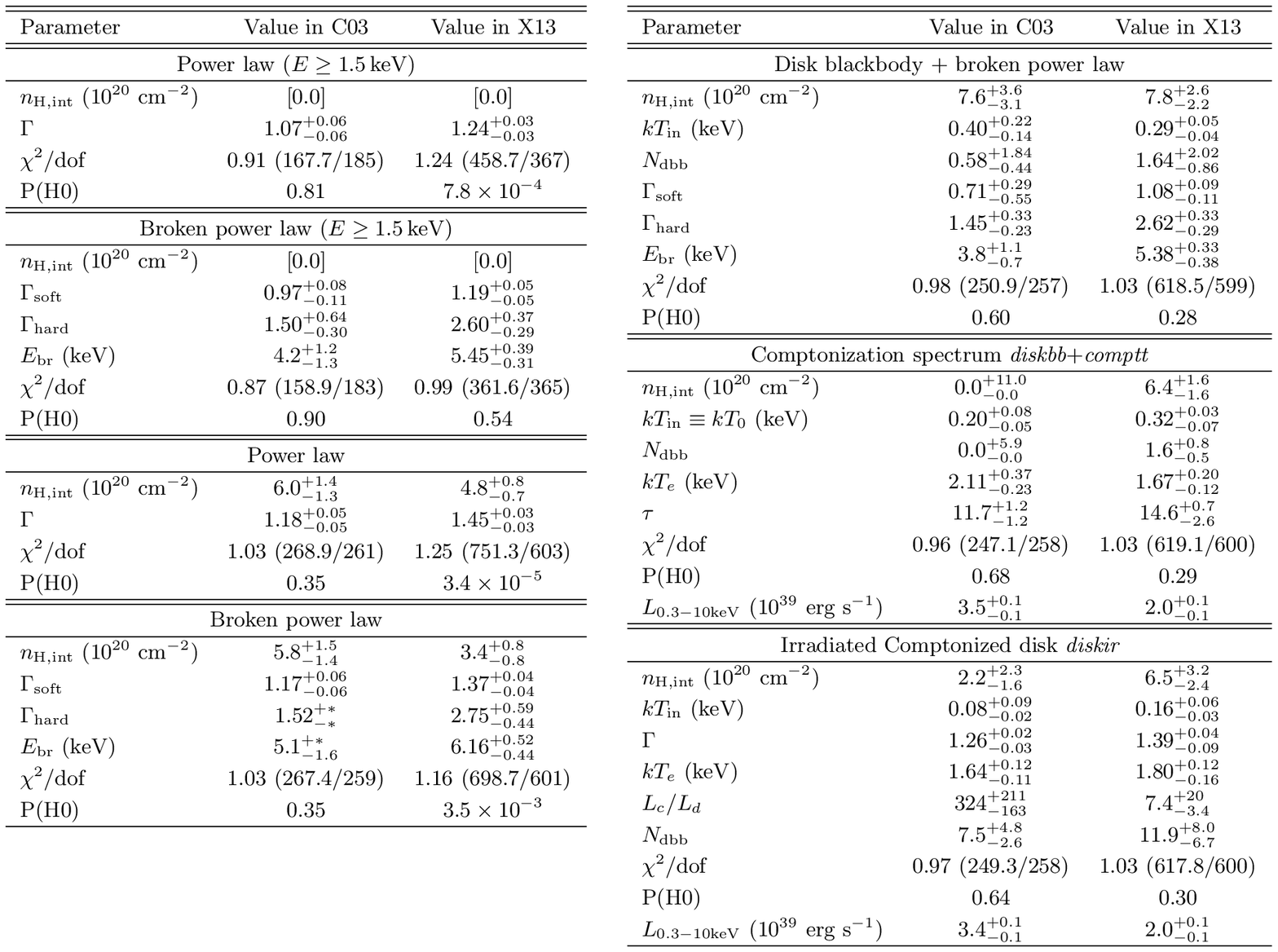},width=16cm,bblly=360pt,clip=true}
\label{tabext:xrayfits}

\clearpage

\begin{figure}[H]
\centerline{\psfig{file=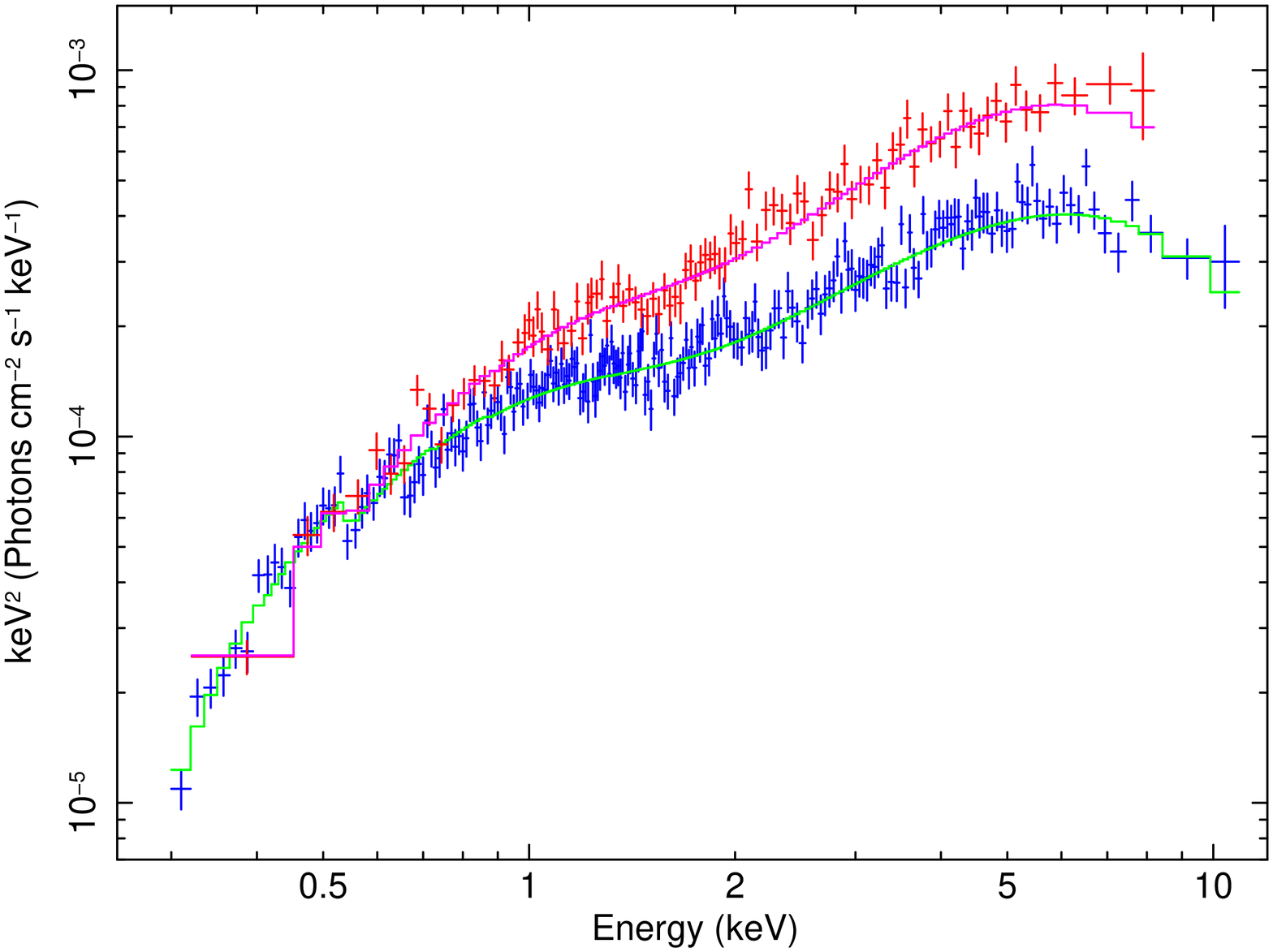,width=15cm,clip=false,bbllx=70pt,bblly=10pt,bburx=580pt,bbury=720pt,angle=-90}}
\caption[]{}
\label{figext:ChandraXMMSpec}
\end{figure}

\begin{figure}[H]
\centerline{\psfig{file=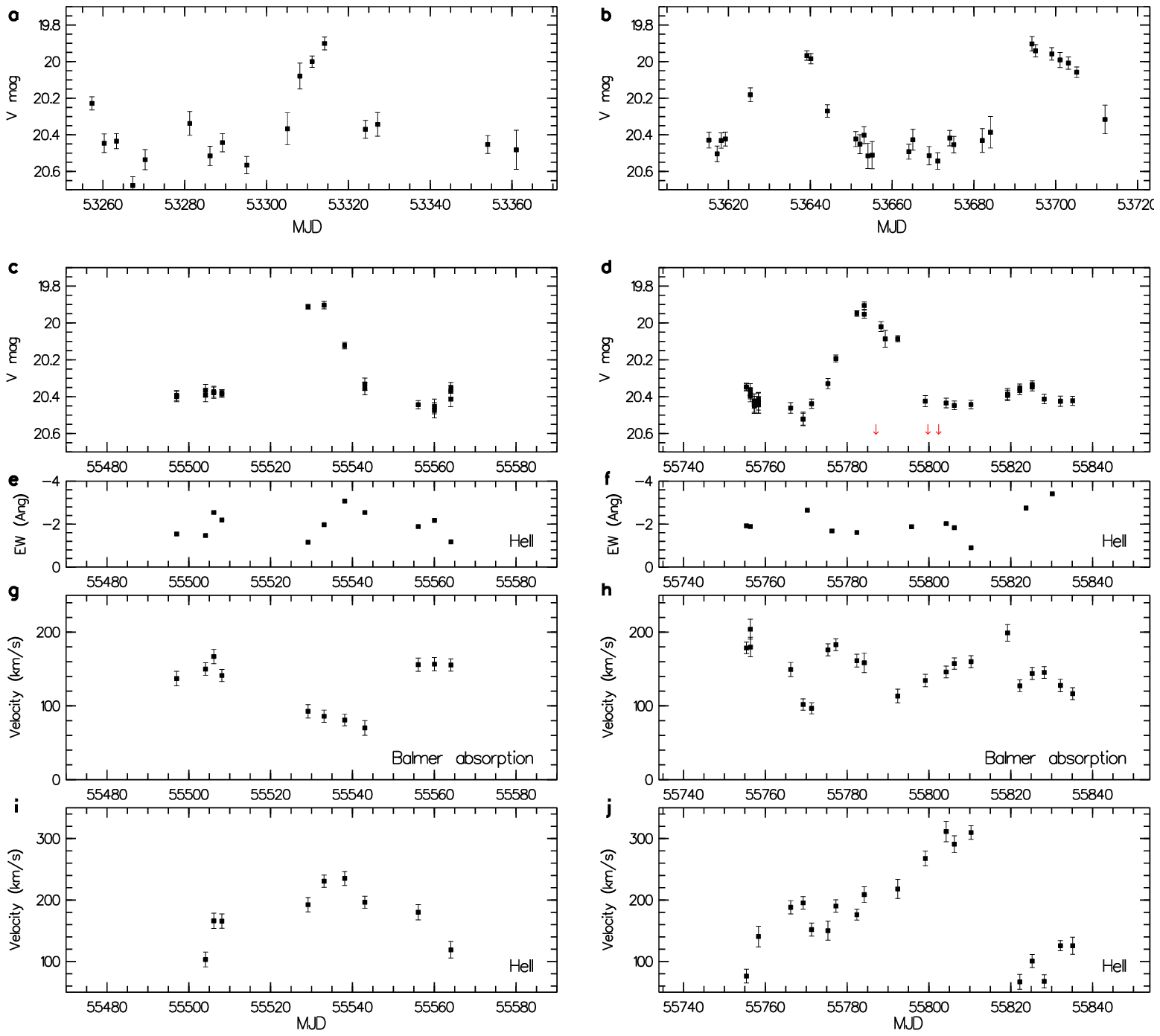,width=15cm,clip=false,bbllx=20pt,bblly=340pt,bburx=520pt,bbury=760pt,angle=0}}
\caption[]{}
\label{figext:opticaldata}
\end{figure}

\begin{figure}[H]
\centerline{\psfig{file=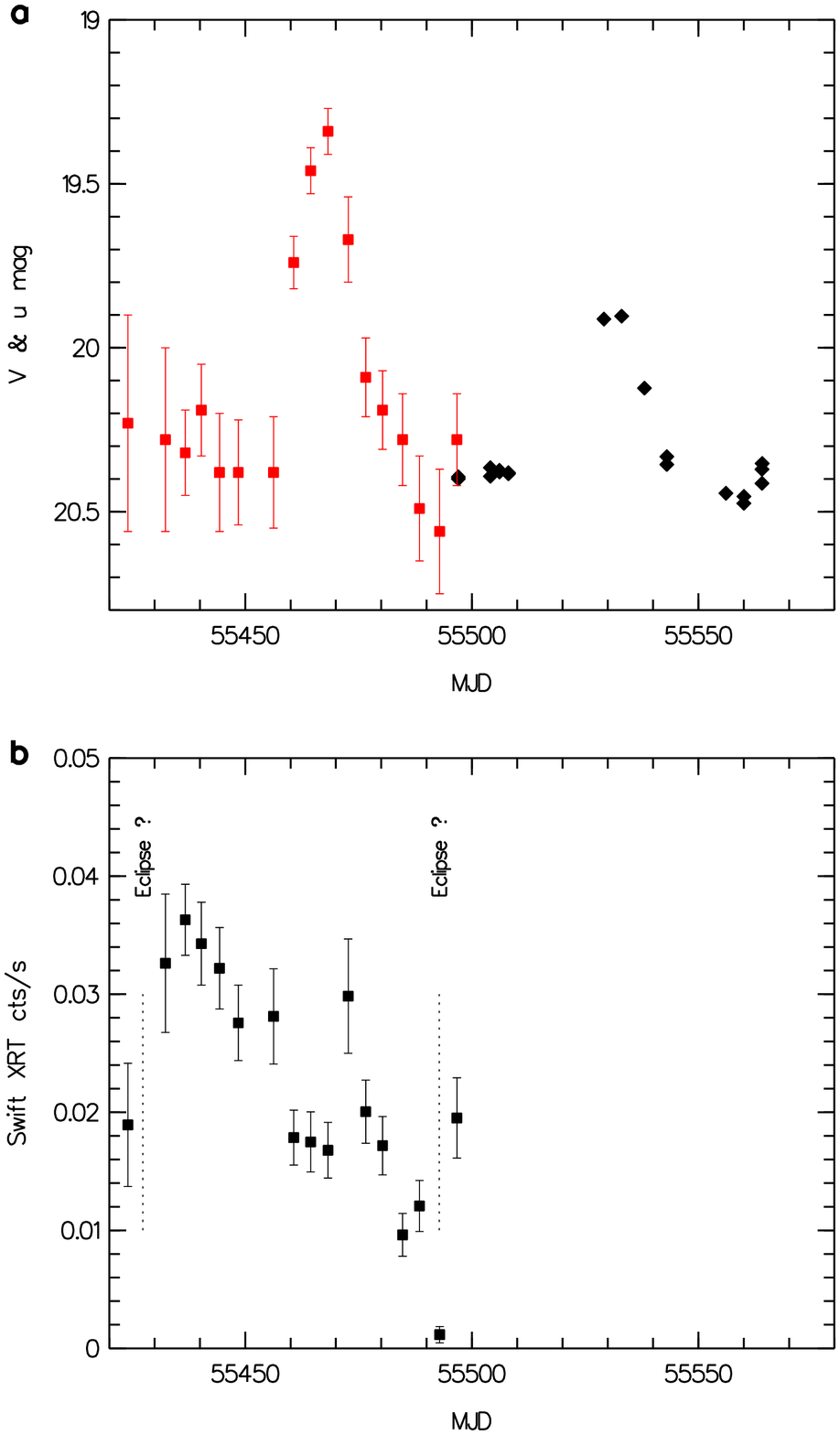,width=10cm,angle=0}}
\caption[]{}
\label{figext:Swift}
\end{figure}

\begin{figure}[H]
\centerline{\psfig{file=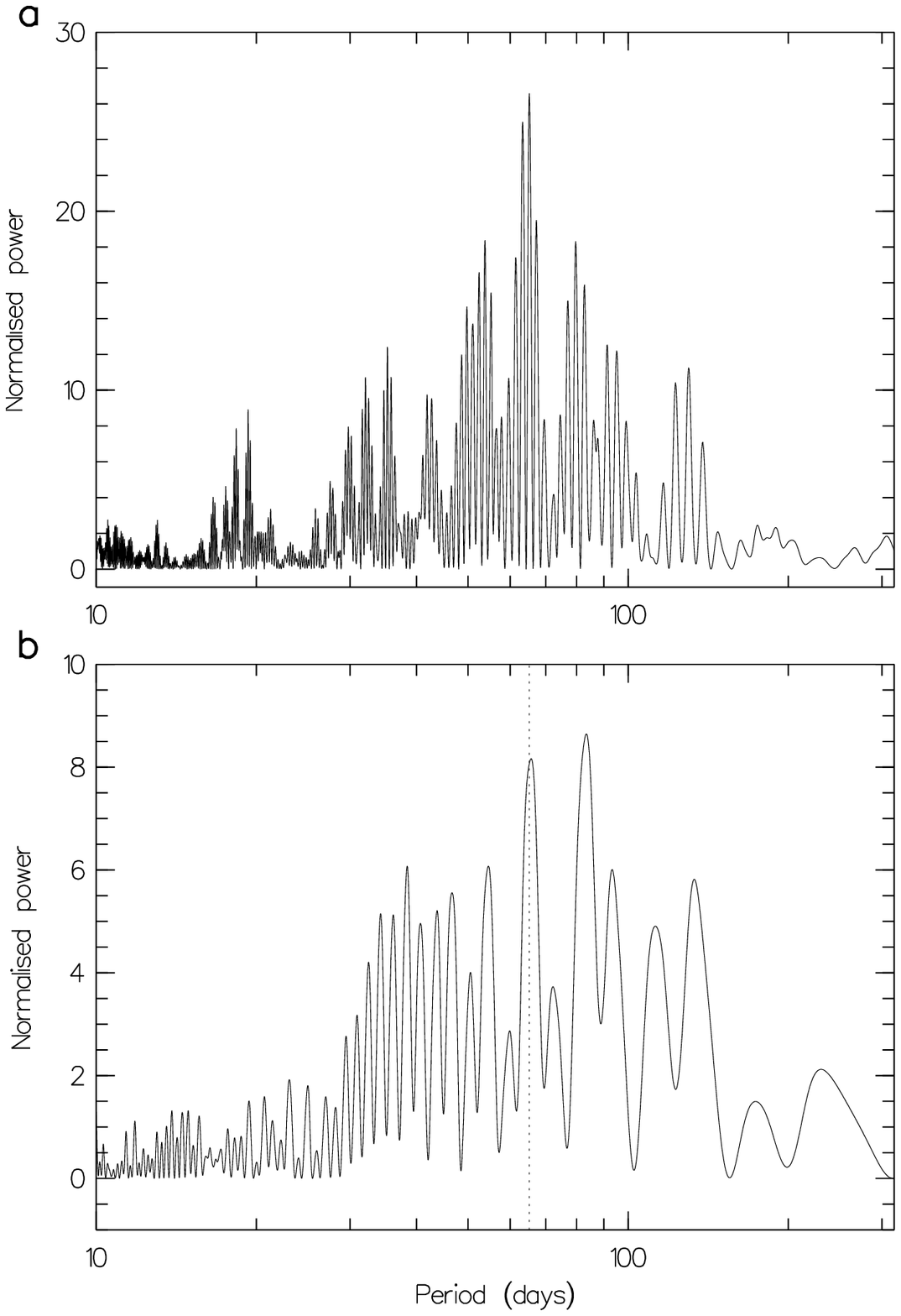,width=15cm,clip=false,bbllx=50pt,bblly=120pt,bburx=520pt,bbury=760pt,angle=0}}
\caption[]{}
\label{figext:power_Vheii}
\end{figure}

\begin{figure}[H]
\centerline{\psfig{file=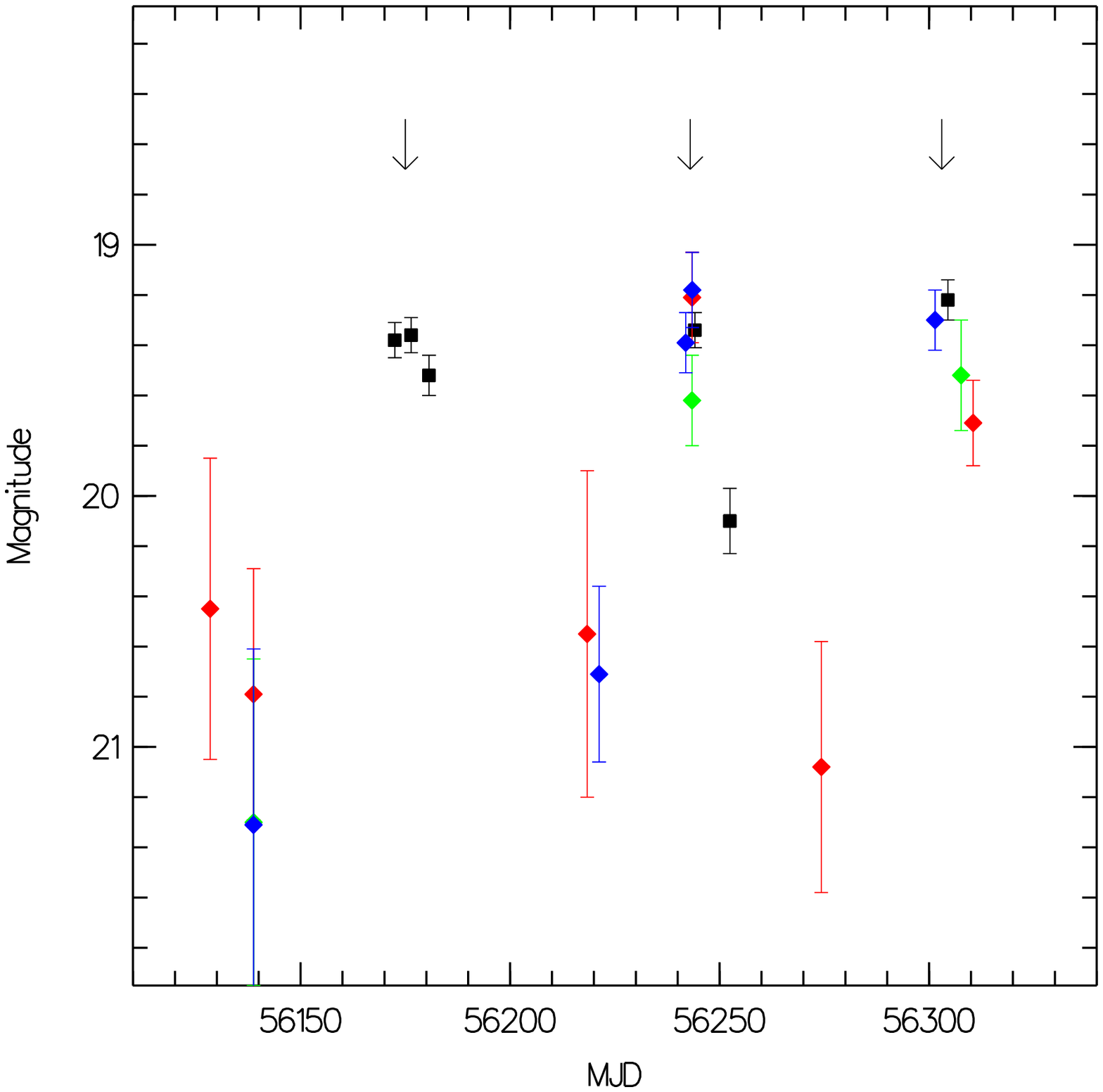,width=5in,bbllx=30pt,bblly=70pt,bburx=580pt,bbury=630pt,clip=false,angle=-90}}
\caption[]{}
\label{figext:UVOTmaxima}
\end{figure}

\begin{figure}[H]
\vskip 1cm
\centerline{\psfig{file=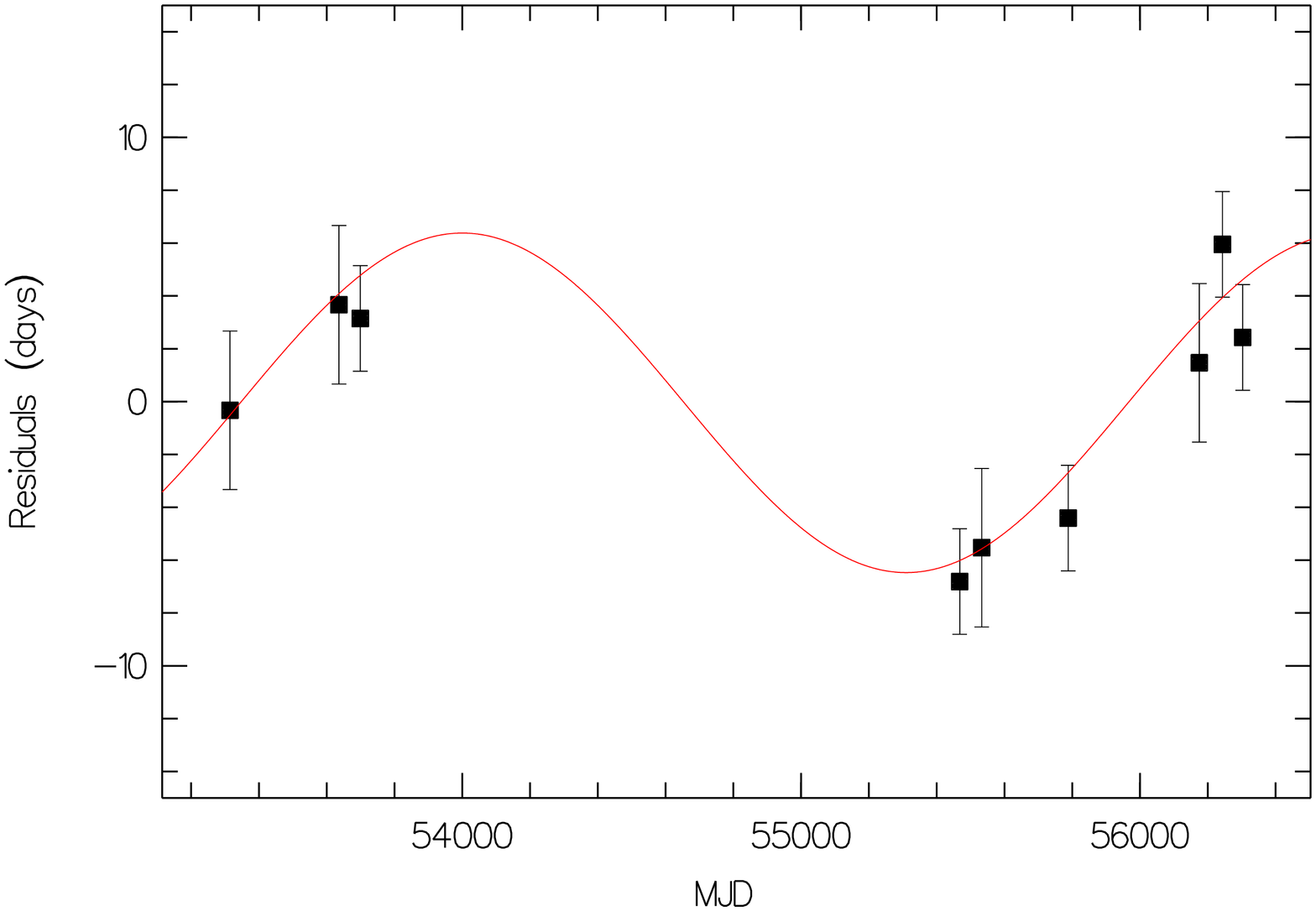,width=12cm,angle=-90}}
\caption[]{}
\label{figext:PeriodFitResiduals}
\end{figure}

\begin{figure}[H]
\centerline{\psfig{file=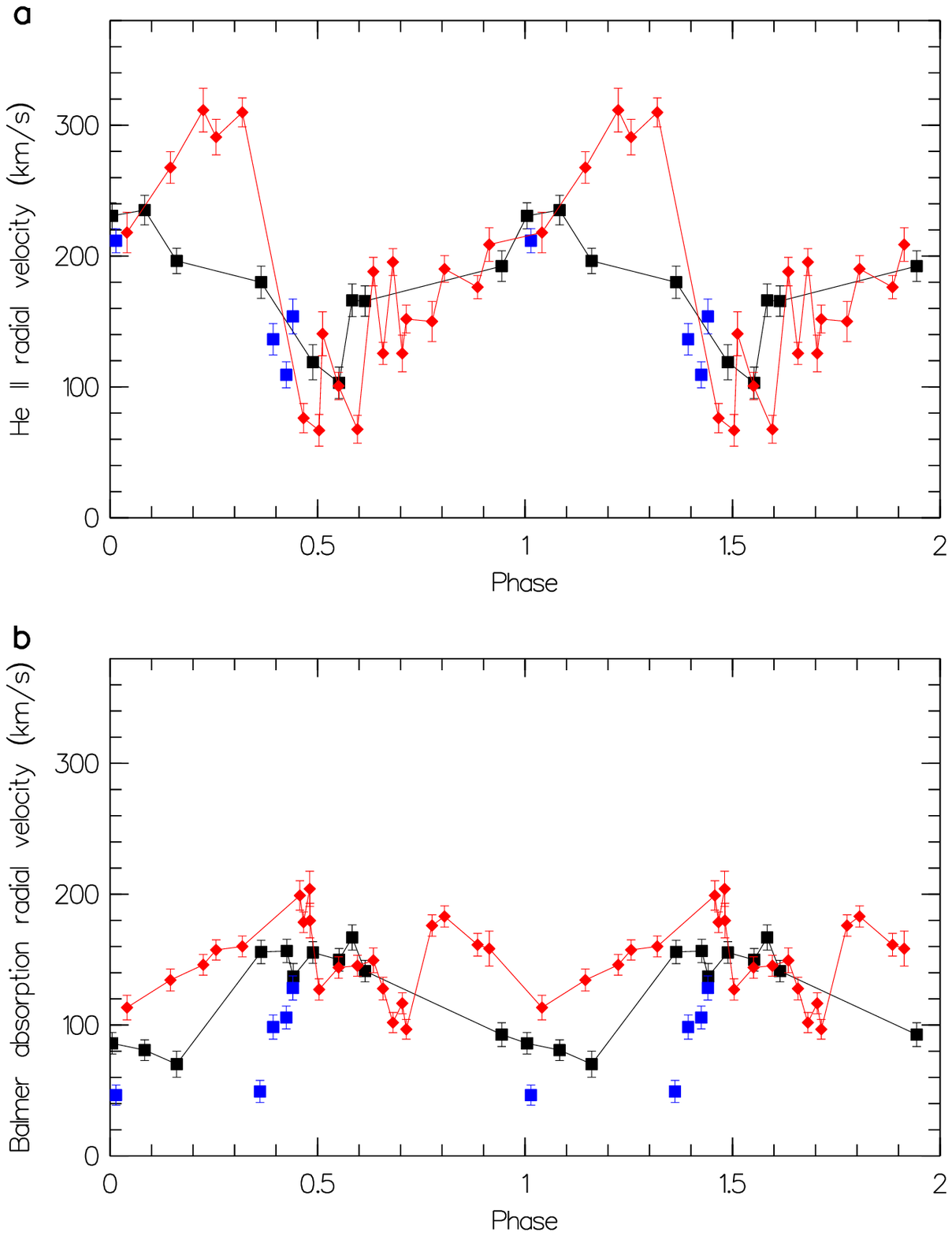,width=15cm,clip=false,bbllx=50pt,bblly=140pt,bburx=520pt,bbury=760pt,angle=0}}
\caption[]{}
\label{figext:foldedHeII}
\end{figure}

\begin{figure}[H]
\centerline{\psfig{file=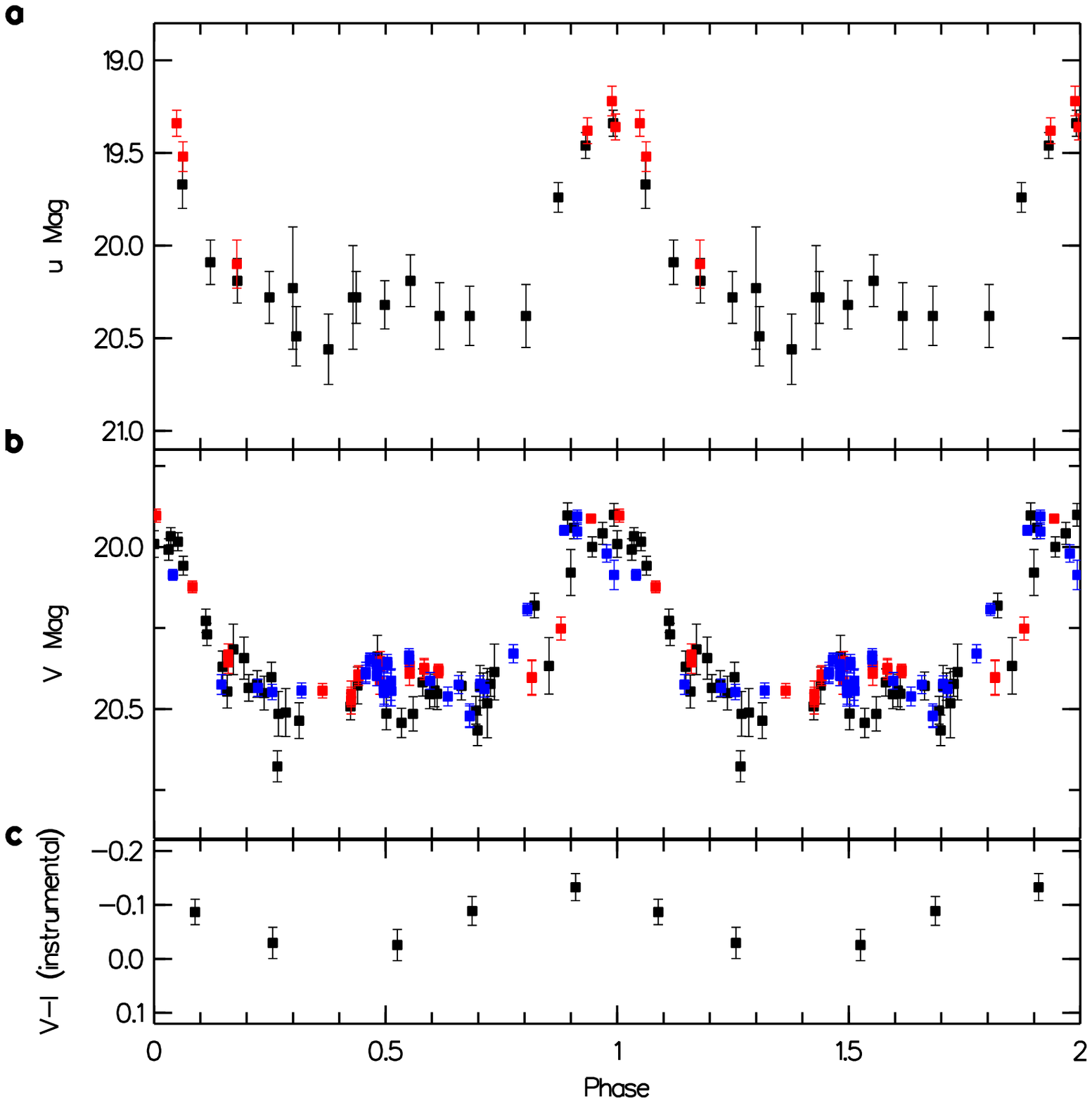,bbllx=30pt,bblly=230pt,bburx=550pt,bbury=740pt,width=16cm,angle=0,clip=true}}
\caption[]{}

\label{figext:foldedLC}
\end{figure}

\begin{figure}[H]
\centerline{\psfig{file=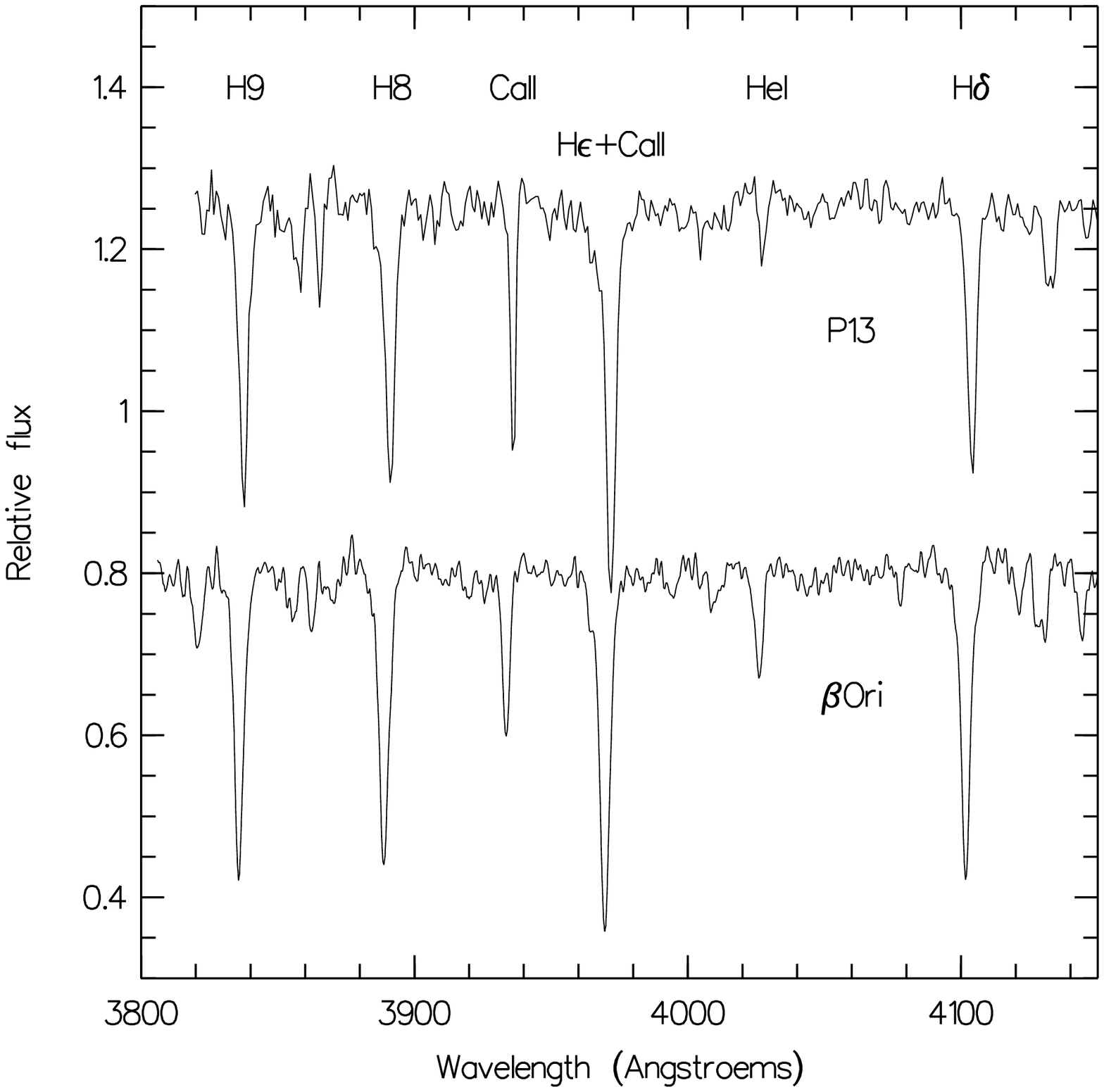,width=13cm,angle=-90,clip=true}}
\caption[]{}
\label{figext:EWcomparison}
\end{figure}

\end{document}